\documentclass{article}

\usepackage{PRIMEarxiv}

\usepackage[utf8]{inputenc} 
\usepackage[T1]{fontenc}    
\usepackage{hyperref}       
\usepackage{url}            
\usepackage{microtype}      
\usepackage{graphicx}       
\graphicspath{{figures/}}   

\usepackage{tabularx,booktabs,array,amsmath}
\usepackage{multirow}
\usepackage{bbm}
\usepackage{enumitem}
\usepackage[table]{xcolor}
\usepackage{colortbl}
\usepackage{subcaption}
\usepackage{makecell}
\newcolumntype{L}{>{\raggedright\arraybackslash}X}
\newcolumntype{P}[1]{>{\raggedright\arraybackslash}p{#1}}

\usepackage{amssymb}
\usepackage{mathtools}
\usepackage{amsthm}
\usepackage[capitalize,noabbrev]{cleveref}

\usepackage{natbib}
\theoremstyle{plain}
\newtheorem{theorem}{Theorem}[section]

\theoremstyle{definition}
\newtheorem{definition}[theorem]{Definition}

\theoremstyle{remark}

\definecolor{lightgray}{gray}{0.93}

\title{LogiC-Diff: Embedding Security Properties Into AI-Enabled Cyber-Physical Systems}

\author{
  Ziyan An \\
  Department of Computer Science \\
  Vanderbilt University \\
  Nashville, TN \\
  \texttt{ziyan.an@vanderbilt.edu} \\
  \And
  John Stankovic \\
  Department of Computer Science \\
  University of Virginia \\
  Charlottesville, VA \\
  \texttt{jas9f@virginia.edu} \\
  \And
  Meiyi Ma \\
  Department of Computer Science \\
  Vanderbilt University \\
  Nashville, TN \\
  \texttt{meiyi.ma@vanderbilt.edu} \\
}

\begin{document}
\maketitle

\begin{abstract}
AI-enabled Cyber-Physical Systems (CPS) are highly vulnerable to adversarial and anomalous inputs, where small perturbations can induce cascading errors and unsafe control actions.
Existing approaches, such as rule-based filtering, training-time regularization, or diffusion-based reconstruction, either operate outside the model or lack mechanisms to incorporate formal security specifications into the prediction process.
In this paper, we take the first step toward embedding security properties directly into AI-enabled CPS, enabling predictive models to enforce system-level constraints during inference rather than relying on external defenses.
We introduce a logic-conditioned bi-stage diffusion framework that integrates Signal Temporal Logic (STL) specifications into forecasting. STL serves as a first-class conditioning signal that guides both an input repair stage and an output refinement stage, allowing the model to jointly mitigate adversarial perturbations and enforce desired temporal behaviors to satisfy security-critical properties.
We evaluate our approach on two real-world multivariate CPS forecasting datasets under a diverse set of physical sensor and cyber attacks. Across sensor faults, gradient-based attacks, adaptive attacks, and varying attack strengths, our method consistently improves robustness and specification compliance, degrades more gracefully as attack strength increases, and generalizes better to unseen attacks. Ablation studies on specification coverage and quality further show that embedding logical security properties yields gains unattainable by reconstruction-based methods alone, highlighting a new direction for integrating formal methods with generative models in secure CPS.
\end{abstract}


\section{Introduction}
AI-enabled cyber-physical systems (CPS), such as smart healthcare, smart cities, and autonomous mobility, rely heavily on deep learning-based forecasting models to support real-time monitoring and control~\citep{yao2019learning,wu2021spatiotemporal,sefati2023meet,mengash2023deep}. However, these systems operate in high-stakes environments where failures can lead to unsafe or even catastrophic outcomes. A key vulnerability arises from data-level adversarial or anomalous inputs: even small $\ell_2$-bounded perturbations can distort sensor signals, propagate through forecasting pipelines, and ultimately trigger unsafe downstream control actions~\citep{liu2022practical,wu2022small,duo2022survey}. As a result, ensuring the robustness of forecasting models is critical to the safe deployment of AI-enabled CPS.

At the core of this challenge is a fundamental limitation of current learning-based forecasting models. These models are largely data-driven and unconstrained, with no built-in guarantees of safety, temporal consistency, or adherence to system-level requirements. When inputs are corrupted, prediction errors can propagate and amplify across the pipeline, leading to cascading failures. This exposes a key gap: current forecasting models lack mechanisms to enforce system-level security properties during prediction. In this paper, we ask a central question: \textit{can system-critical security properties be embedded into learning-enabled CPS by design, so that models remain robust under adversarial attacks?}

Existing defenses for CPS forecasting fall into two broad categories. The first focuses on improving robustness through learning-based reconstruction methods~\citep{cai2021generative,salman2020denoised,wu2023defending}. Early approaches such as Defense-GAN~\citep{samangouei2018defense} detect or repair adversarial inputs by exploiting distributional discrepancies, while more recent diffusion-based models restore samples via learned reverse processes~\citep{sohl2015deep,ho2020denoising}. Although these methods achieve strong empirical robustness, they rely solely on distributional reconstruction and lack an explicit notion of what constitutes safe or acceptable system behavior. As a result, they cannot enforce temporal or logical constraints on predictions.

A complementary line of work incorporates rule-based properties to improve alignment with system requirements under nominal conditions~\citep{ma2020stlnet,hu2016harnessing,an2024formal}. These approaches encode semantic constraints—such as temporal ordering, persistence, and cross-feature consistency—using formal languages like Linear Temporal Logic~\citep{pnueli1977temporal}, Metric Temporal Logic~\citep{koymans1990specifying}, and Signal Temporal Logic (STL)~\citep{maler2004stl}. In practice, such properties are applied as training-time regularizers or post-hoc monitoring and filtering mechanisms~\citep{swaroop2024comprehensive,dmitry2022formal,bartocci2018specification,yang2022improving}. However, these approaches are not designed as security mechanisms: they operate outside the predictive process, are reactive rather than preventative, and do not mitigate adversarial perturbations during inference. This limitation is exacerbated under structured perturbations, such as gradient-based attacks that shift inputs across fragile decision boundaries: once temporal dynamics are distorted before prediction, the model may apply a qualitatively different mapping to the corrupted data~\citep{hein2017formal,wesegoadversary}. Indeed, post-hoc correction can enforce certain output constraints, but it cannot reliably recover information lost upstream, and the resulting forecast may still fail to reflect the true system dynamics.

In this work, we take a step toward embedding security properties directly into the forecasting process. We propose a logic-conditioned bi-stage diffusion framework that integrates formal specifications into model inference. Specifically, we use Signal Temporal Logic (STL) to express system-critical properties and incorporate them as conditioning signals within a diffusion-based repair pipeline. This enables a new paradigm of specification-aware prediction, where forecasts are actively shaped to satisfy security constraints while maintaining robustness.

Our work investigates repositioning prior secure system specifications, expressed as temporal logic, into two stages of the diffusion-based repair process, thereby enabling both pre-hoc and post-hoc defense. Specifically, we condition the diffusion-based repair model on a logic-informed signal obtained by projecting the data manifold, at different stages of the framework, onto the $\ell_1$ minimum-distance $\varphi$-feasible set. We then modulate the diffusion prediction with a learnable gating coefficient that balances the diffusion reconstruction and the projected signal in the final output. The framework is trained end-to-end under loss terms that cover each of our three defense desideratum. 

We evaluate our method on two real-world multivariate time-series CPS forecasting datasets, including a smart city dataset spanning Caltrans~\cite{chen2002freeway} and air quality and a UAV dataset~\cite{keipour2021alfa}. The framework is evaluated against a comprehensive set of black-box sensor attacks and faults, gradient-based white-box attacks, adaptive and query-based attacks, sparse indirect attacks, and varying attack strengths. We further study cross-attack generalization, cross-domain transfer, and sensitivity to specification coverage and quality. Compared with a broad set of baselines spanning adversarial training, adversarial purification, randomized smoothing, and logic-regularized learning, our approach consistently reduces forecasting error and improves specification compliance and generalization across a range of attack strengths. Our ablation results further show that these gains depend strongly on the guidance provided by the embedded logic specifications. 

To our knowledge, this is the first work to embed formal security properties directly into CPS forecasting via logic-conditioned generative modeling.
\textbf{Contributions.} (1) We formulate secure CPS forecasting as embedding STL-based specifications into learning-based prediction. (2) We propose a logic-conditioned bi-stage diffusion framework for input repair and output refinement. (3) We develop an STL-guided conditioning mechanism that enables specification-aware generative reconstruction under adversarial settings. (4) We conduct a comprehensive evaluation across two CPS domains and a diverse set of threat models, including physical sensor attacks, gradient-based and adaptive cyber attacks, varying attack strengths, sparse indirect attacks, and unseen attack settings. The results show that \textsc{LogiC-Diff} consistently improves robustness and specification compliance while demonstrating stronger generalization across attacks and domains. 
\section{Background and Problem Statement}

We introduce the key preliminaries underlying our approach, including DDPM models~\citep{ho2020denoising} for generative repair, Signal Temporal Logic (STL)~\citep{maler2004stl} for formal specification, and the CPS security setting considered in this work.

\paragraph{Background: Diffusion Models for Data Repair.}
Diffusion models~\citep{sohl2015deep,song2020score} learn a data distribution $p(x)$ through gradual noising and denoising. In the forward process, clean data $x_0$ are corrupted by Gaussian noise over $\mathrm{N}$ steps until $x_\mathrm{N} \sim \mathcal{N}(0,I)$. We build on DDPMs~\citep{ho2020denoising} as the basis of our input repair mechanism, where a reverse diffusion network $\epsilon_\theta(x_n,n)$ predicts the injected noise and denoises corrupted inputs back toward the clean data manifold.

In forecasting settings with adversarial purification, a potentially corrupted input $x^\ast = x+\delta$ is treated as a noisy observation of an underlying clean sample $x$ and is first injected with Gaussian noise during the forward diffusion step. In the reversing step, a denoiser $\epsilon_\theta$ parameterized by $\theta$ guides the learned reverse process to produce the reconstructed data $\tilde{x}=\Phi_\theta(x^\ast)$, where $\Phi_\theta$ denotes the forward-reverse denosing process. In the subsequent CPS forecasting step, a forecasting neural network $F_\omega$ is applied to $\tilde{x}$, with the goal that its prediction approximates the clean-input forecast $F_\omega(x)$. 

\paragraph{Background: Signal Temporal Logic.}\label{sec:prelim-stl}


STL is a real-valued, time-bounded formalism for specifying temporal properties over continuous signals. Given time-series data and a CPS specification, STL supports monitoring through qualitative semantics that describe Boolean satisfaction and quantitative semantics that measure the degree of satisfaction or violation~\citep{donze2010robust}. In this work, we use STL to formalize ``secure'' system behaviors, including temporal requirements such as event ordering and persistence, as well as relationships between multiple signal traces over specified time intervals.

Consider an $[a,b]$-bounded time interval and an atomic predicate of the form $\mu \equiv f(x_t) \ge c$, where $f:\mathbb{R}^d \rightarrow \mathbb{R}$ is a valuation function and $c \in \mathbb{R}$ is a threshold. A formula $\varphi$ is defined inductively as $\varphi := \mu \mid \neg \varphi \mid \varphi_1 \wedge \varphi_2 \mid \varphi_1 \mathcal{U}_{[a,b]} \varphi_2 \mid \square_{[a,b]} \varphi \mid \lozenge_{[a,b]} \varphi.$
Here, the temporal operators include the ``always'' operator $\square_{[a,b]}$, which requires that a time series satisfy $\varphi$ at everywhere within the bounded interval; the ``eventually'' operator $\lozenge_{[a,b]}$, which requires that $\varphi$ hold at least once within the bounded interval; and the ``until'' operator $\mathcal{U}_{[a,b]}$, which requires that $\varphi_1$ hold until $\varphi_2$ is true within the bounded interval. 
The STL quantitative semantics function $\rho_\varphi(x, t) \in \mathbb{R}$ measures the signed margin by which a signal satisfies or violates $\varphi$ at time $t$. We refer the reader to App.~\ref{sec:stl-semantics} for the full definition of STL semantics and additional examples of logic rules. 

\paragraph{Threat Model.}
We consider a comprehensive set of runtime attacks for CPS spanning two attack families, both of them aim to corrupt the forecasting model's output and adversely affect downstream system performance. The first family consists of black-box CPS \textit{physical} system attacks, which we refer to as sensor attacks~\citep{yu2023survey}. In this setting, the attacker has access to the physical or communication layer and can compromise sensor data by freezing, replaying, or spoofing sensor readings, as well as by applying randomized mixtures of these attack types~\citep{duo2022survey}. 

The second family consists of gradient-based white-box attacks to the \textit{cyber} system, bounded by an $\ell_2$ budget $\|\delta\|_2 \le \varepsilon$, where $\delta$ is the added perturbation. These attacks leverage model gradients to generate the worst-case adversarial input within the allowed budget, which is typically formulated as $x^\ast \in \arg\max_{x'} \mathcal{L}(x',y;\omega) \; \text{s.t.} \|x' - x\|_2 \le \varepsilon,$
where $\mathcal{L}$ is the model's training loss, $(x,y)$ is a clean input-output pair, and $\omega$ denotes the model parameters~\citep{goodfellow2014explaining}. We consider specific attacks including PGD~\citep{madry2018towards}, MIM~\citep{dong2018boosting}, CW~\citep{carlini2017towards}, and AutoAttack~\citep{croce2020reliable}.

\paragraph{Defense Goal.}\label{sec:defense-goal}
Given a potentially attacked input trajectory $x^\ast$, our goal is to produce a repaired trajectory $\tilde{x}$ and subsequently a final forecast $\tilde{y}$ that are both accurate and semantically valid. Specifically, the defense should produce:
(i) a repaired input trajectory $\tilde{x}$ that remains close to the clean input $x$ and satisfies an input-side STL specification $\varphi$, and (ii) a final forecast $\tilde{y}$ that approximates the clean-input forecast $F_\omega(x)$ and satisfies an output-side STL specification $\varphi'$. 
Formally, we have: 
\begin{equation*}
F_\omega(\tilde{x}) \approx F_\omega(x), \quad
\tilde{x} \in \arg\max_{\bar{x} \in \mathcal{N}(x^\ast)} \rho_\varphi(\bar{x}, t), \quad
\tilde{y} \in \arg\max_{\bar{y} \in \mathcal{N}(F_\omega(\tilde{x}))} \rho_{\varphi'}(\bar{y}, t), 
\end{equation*} 
where $\mathcal{N}(\cdot)$ denotes a projected neighborhood of the given point.
Together, these objectives require a defense that produces outputs both likely under the clean data distribution and consistent with system-level security specifications. We describe how our bi-stage diffusion framework operationalizes these goals in the next section. 
\begin{figure*}[t]
    \centering
    \includegraphics[width=0.9\linewidth]{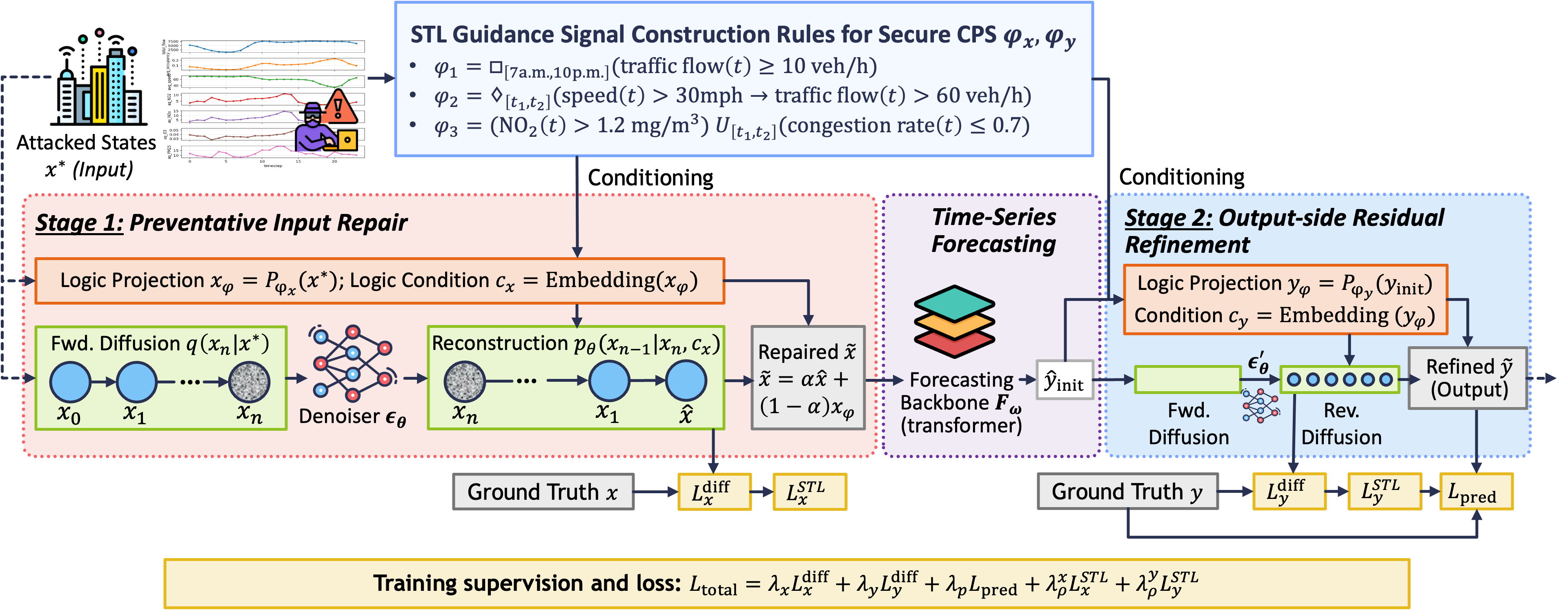}
    \caption{Overview of LogiC-Diff. Given a potentially attacked multivariate CPS time-series input, LogiC-Diff first constructs STL-guided signals from CPS security specifications. The framework then performs a repair and a refinement stage on input-side and an output-side, respectively. In both stages, the diffusion estimate is blended with the projection to encourage both data-plausible reconstruction and specification-consistent behavior. The framework is trained end-to-end.}
    \label{fig:framework-all}
    \vspace{-1em}
\end{figure*}

\section{Method: LogiC-Diff} 
We present our logic-guided bi-stage diffusion defense for secure CPS forecasting. At a high level, the method first projects STL specifications into a logic-consistent guidance signal, and then uses this signal to nudge the reconstruction process at two repair stages: an input-stage diffusion that preventatively reconstructs the attacked trajectory before forecasting, and an output-stage diffusion that refines the forecast residual afterward. We refer to this pipeline as \textsc{LogiC-Diff}, short for STL-conditioned diffusion reconstruction, and use $\Phi_\theta$ to denote the full end-to-end pipeline. Figure~\ref{fig:framework-all} depicts the overall workflow of our framework. 

\subsection{STL Guidance Signal}\label{sec:stl-robust}
In this section, we formalize how STL rules support defense against security attacks. We introduce CPS-oriented spatiotemporal logic formulas that encode safe global behavior and are also relevant to downstream control, then describe how an attacked signal is projected into a guidance signal. This signal is later used as the basis for an embedding to condition the diffusion model and as a reference trajectory in the final interpolation step, encouraging specification-compliant repair. 

\paragraph{Security Guidance: STL Rules.} 
In Section~\ref{sec:prelim-stl}, we introduced the formal definition of STL and its temporal operators. In our setting, an STL rule takes a form such as $\varphi := \square_{[t_1,t_2]}(x_t \geq c)$, where the temporal operator $\square_{[t_1,t_2]}$ may be replaced by $\lozenge_{[t_1,t_2]}$ depending on the desired requirement, and $\mu := (x_t \geq c)$ denotes the atomic predicate that the temporal requirement is based on. 

Specifically, we define the following set of spatiotemporal logic formulas to support the security requirements of CPS operation. Let $x^{(j)}_t$ denote the value of the $j$-th feature at time $t$ in the forecasting dataset. We then apply the following coarse categories of logic rules, summarized in Table~\ref{tab:stl-rules}. 

\begin{table*}[t]
\centering
\caption{STL-based CPS specifications. Each rule is parameterized by features and a time interval $[t_1, t_2]$ and encodes a physically meaningful constraint on feature trajectories.} 
\label{tab:stl-rules}
\small
\resizebox{\textwidth}{!}{%
\begin{tabular}{@{}l | p{0.618\textwidth}@{}}
\toprule
\multicolumn{2}{@{}l}{\textbf{STL-based Guidance Signal Construction Rules for Secure CPS}} \\
\midrule
$\varphi_1 := \square_{[t_1,t_2]}\bigl((x^{(j)}_t \ge c_1)\wedge \neg(x^{(j)}_t \geq c_2)\bigr)$
  & \texttt{Stability}$(x^{(j)},[t_1,t_2])$ enforces physically valid ranges and prevents false high or low values outside feasible limits over time. \\
\midrule
$\varphi_2 := \square_{[t_1,t_2]}\Bigl(\neg\bigl((x^{(j)}_t \geq c_1) \wedge \neg(x^{(j+1)}_t \geq c_2)\bigr)\Bigr)$ & \texttt{Associativity}$(x^{(j)},x^{(j+1)},[t_1,t_2])$ encapsulates the correlational associativity across features over time. \\
\midrule
$\varphi_3 := \neg(x^{(j)}_t \geq c_1)\,\mathcal{U}_{[t_1,t_2]}\,(x^{(j+1)}_t \geq c_2)$
  & \texttt{Propagation}$(x^{(j)},x^{(j+1)},[t_1,t_2])$ strengthens upstream and downstream system behavior between two sensor locations or across temporally related events. \\
\midrule
$\varphi_4 := \square_{[t_1,t_2]}\Bigl(\neg\bigl(|\bar{x}^{(j)}_{[t-w_1,t]} - \bar{x}^{(j)}_{[t-w_1-w_2,t-w_1]}| \geq c_1\bigr)\Bigr)$ & \texttt{Smoothness}$(x^{(j)},[t_1,t_2],w_1,w_2)$ bounds the change in mean between adjacent windows of widths $w_1$ and $w_2$, detecting persistent changes in feature behavior. \\
\midrule
$\varphi_5 := \square_{[t_1,t_2]}\Bigl(\neg\bigl((x^{(j)}_t \geq c_1) \wedge \neg\lozenge_{[d_1,d_2]}(x^{(j+1)}_t \geq c_2)\bigr)\Bigr)$
  & \texttt{Latency}$(x^{(j)},x^{(j+1)},[t_1,t_2],[d_1,d_2])$ enforces that whenever feature $x^{(j)}$ reaches level $c_1$, feature $x^{(j+1)}$ reaches level $c_2$ at some time within the future window $[d_1, d_2]$. \\
\midrule
$\varphi_6 := \lozenge_{[t_1,t_2]}\,\square_{[0,d]}\bigl((x^{(j)}_t \geq c_1) \wedge \neg(x^{(j)}_t \geq c_2)\bigr)$
  & \texttt{Recovery}$(x^{(j)},[t_1,t_2],d)$ requires the feature to eventually enter and remain within a safe operational band for a duration of at least $d$ time units. \\
\bottomrule
\end{tabular}
}
\vspace{-1em}
\end{table*}


More specific examples specifications are listed under App.~\ref{sec:apdx-spec}. In practice, multiple STL rules can be combined through logical conjunction. We use $\varphi$ to denote the set of rules applied for input reconstruction, and $\varphi'$ to denote the set of rules applied for output-side residual repair and reconstruction. In many applications, these rules can be specified directly by system engineers; alternatively, several existing tools like \textsc{TeLEx} and \textsc{Quivr} can infer or construct such rules from sample data or query specifications~\citep{jha2019telex, mell2023synthesizing}. 

\paragraph{Guidance Signal Construction.} We have established several categories of rules that define secure system behavior. We now describe how a logic specification is grounded into a guidance signal for diffusion-based reconstruction. Following prior work~\citep{ma2020stlnet}, we project the attacked signal $x^\ast$ onto the $\varphi$-feasible set, yielding a logic-consistent reference trajectory. We define the projection operator as $P_{\varphi}(x),$ such that it yields the closest satisfying trace in $\ell_1$ distance. 

The projection treats logical and temporal operators compositionally through the quantitative STL semantics. For an atomic predicate $\mu = (x_t \geq c)$, the projection clamps each violating time point upward to the threshold, $x'_t = \max(x^\ast_t, c)$, while the satisfying points are unchanged. For $\square_{[a,b]}\mu$ operator, which requires $\mu$ to hold at every $t \in [a,b]$, we apply this clamping at every time point in the interval, $x'_t = \max(x^\ast_t, c)$ for all $t \in [a,b]$. For the $\lozenge_{[a,b]}\mu$ operator, which requires $\mu$ to hold at some $t \in [a,b]$, we identify the time point already closest to satisfaction, $\hat{t} = \arg\min_{t \in [a,b]}\mid x^\ast_t-c\mid$, and clamp only that point, $x'_{\hat{t}} = \max(x^\ast_{\hat{t}}, c)$. For the $\mu_1 \mathcal{U}_{[a,b]} \mu_2$ operator, which requires $\mu_1$ to hold up to some time $\hat{t} \in [a,b]$ at which $\mu_2$ holds, we select $\hat{t}$ to minimize the joint projection distance, then enforce $\mu_1$ on $[a, \hat{t}]$ and $\mu_2$ at $\hat{t}$ via the projections above. 

For Boolean operators, the conjunction $\varphi_1 \wedge \varphi_2$ requires both subformulas to hold and is projected by sequentially applying the projection of each conjunct, $P_{\varphi_1 \wedge \varphi_2}(x^\ast) = P_{\varphi_2}(P_{\varphi_1}(x^\ast))$; the disjunction $\varphi_1 \vee \varphi_2$ requires either subformula to hold and is projected toward the cheaper of the two, $P_{\varphi_1 \vee \varphi_2}(x^\ast) = \arg\min_{i \in \{1, 2\}} \|P_{\varphi_i}(x^\ast) - x^\ast\|_1$. We obtain the final guidance signal $x_\text{logic} = P_\varphi(x^\ast)$ by iteratively performing operator-level projections over the syntax tree of $\varphi$.

\subsection{Logic-Conditioned Diffusion Reconstruction}
Once the specification-faithful projection of the corrupted input time-series signal $x^\ast$ is obtained, we employ a diffusion reconstruction process to recover $\tilde{x}$. We now describe this process under the discrete-time DDPM formulation. 

\paragraph{Forward Diffusion.} We start from the attacked input $x_0=x^\ast$ and gradually add Gaussian noise according to a $\beta$ schedule over $N$ steps, where $x_1, \dots, x_N$ denote the intermediate latent variables. The forward diffusion process is then formulated as $q(x_n \mid x_{n-1}) = \mathcal{N}(x_n;\sqrt{1-\beta_n}\,x_{n-1},\,\beta_n \mathbf{I}).$
This is equivalent to the closed-form expression $q(x_n \mid x_0)=\mathcal{N}(x_n;\sqrt{\bar{\alpha}_n}\,x_0,\,(1-\bar{\alpha}_n)\mathbf{I}),$
where $\alpha_n = 1-\beta_n$, and $\bar{\alpha}_n$ is the cumulative product of noise retention at step $n$, $\bar{\alpha}_n = \prod_{i=1}^{n}\alpha_i$. 

\paragraph{Logic-Conditioned Reverse Diffusion.} 
For the reverse diffusion process, standard DDPM defines $p_{\theta}(x_{n-1}\mid x_n)=\mathcal{N}(x_{n-1}; \mu_{\theta}(x_n,n), \sigma_{\theta}^2(x_n,n)\mathbf{I}).$
Following this formulation, we incorporate the conditioning signal $c_x$ and obtain $p_{\theta}(x_{n-1}\mid x_n, c_x)=\mathcal{N}(x_{n-1}; \mu_{\theta}(x_n,n,c_x), \sigma_{\theta}^2(x_n,n)\mathbf{I}). $
Here, the mean function $\mu_\theta$ is parameterized through a noise predictor $\epsilon_\theta(x_n,n,c_x)$ and the variance follows the standard DDPM schedule. Rather than iterating this transition over all $N$ steps, we leverage the one-shot $x_0$-estimate implicit in the noise-prediction parameterization:
\begin{equation*}
    \hat{x} = \frac{x_n - \sqrt{1-\bar{\alpha}_n}\,\epsilon_\theta(x_n, n, c_x)}{\sqrt{\bar{\alpha}_n}} = x^\ast+ \frac{\sqrt{1-\bar{\alpha}_n}}{\sqrt{\bar{\alpha}}_n} (\epsilon-\epsilon_\theta). 
\end{equation*}
Here, $\hat{x}$ approximates the conditional mean $\mathbb{E}\,[x_0\mid x_n, c_x]$. To further encourage the repaired sequence to be both data-plausible and specification-consistent, we form the input-side repair as $\tilde{x} = \alpha\,\hat{x} + (1-\alpha)\,x_{\text{logic}}$, where $\alpha \in (0,1)$ is a learnable weight. 

\paragraph{Logic-Conditioned Diffusion Refinement on Outputs.} 
The repaired input $\tilde{x}$ is passed to the forecasting module $F_\omega$ to produce an initial forecast $\hat{y}_\text{init} = F_\omega(\tilde{x})$. To refine $\hat{y}_\text{init}$ and account for residual errors, we then apply a second logic-conditioned diffusion on $\hat{y}_\text{init}$. 

Mirroring the first diffusion behavior, we project $\hat{y}_\text{init}$ onto the $\varphi'$-feasible set to obtain $y_\text{logic}$, and form the conditioning signal $c_y$. The forward process noises $\hat{y}_\text{init}$ over $M$ steps following the same DDPM schedule as above. We use a separate noise predictor $\epsilon_{\theta}'$ and conditioning encoder for the output stage. The conditioned reverse process yields a one-shot estimate of the underlying ground-truth forecast, 
\begin{equation*}
    \hat{y} = \frac{y_m - \sqrt{1-\bar{\alpha}_m}\,\epsilon_{\theta}'(y_m, m, c_y)}{\sqrt{\bar{\alpha}_m}}.
\end{equation*} 
We $\hat{y}$ combine with $y_\text{logic}$ to produce the final refined forecast $\tilde{y} = \alpha'\,\hat{y} + (1-\alpha')\,y_\text{logic}$, where $\alpha' \in (0,1)$ is a learnable weight. 

\subsection{End-to-End Training and Inference}
We train the bi-stage pipeline end-to-end with a joint objective that directly instantiates the defense goal in Section~\ref{sec:defense-goal}. At the input stage, the objective encourages the repaired trajectory $\tilde{x}$ to remain close to the clean signal $x$ while achieving high robustness with respect to the input-side STL specification $\varphi$. At the output stage, it encourages the final forecast $\tilde{y}$ to remain accurate relative to the target forecast and to achieve high robustness with respect to the output-side STL specification $\varphi'$. 

Each desideratum is realized through explicit loss terms. We supervise the induced clean estimates $\hat{x}$ and $\hat{y}$ via the reparameterization trick in~\cite{ho2020denoising}, which maps predicted noise to the corresponding $x_0$-estimate. In practice, the reverse models are still parameterized through noise predictors $\epsilon_\theta$ and $\epsilon_{\theta'}$. On the input side, a reconstruction loss $\mathcal{L}_{\text{diff}}^x = \mathbb{E}\|x-\hat{x}\|_2^2$ pulls the diffusion estimate toward the clean signal, while a smoothed STL penalty $\mathcal{L}_{\text{stl}}^x = \mathbb{E}\,[\mathrm{softplus}(-\tilde{\rho}_{\varphi}(\hat{x},0))]$ discourages residual specification violations. On the output side, a reconstruction loss $\mathcal{L}_{\text{diff}}^y = \mathbb{E}\|y-\hat{y}\|_2^2$ aligns the diffusion estimate with the ground-truth forecast, a smoothed STL penalty $\mathcal{L}_{\text{stl}}^y = \mathbb{E}\,[\mathrm{softplus} (-\tilde{\rho}_{\varphi'}(\hat{y},0))]$ encourages output-side feasibility, and a forecasting loss $\mathcal{L}_{\text{pred}} = \mathbb{E}\|y-\tilde{y}\|_2^2$ ensures that the final blended forecast remains accurate. Here, $\tilde{\rho}$ denotes the smooth surrogate of STL robustness~\citep{leung2023backpropagation}, and the softplus function provides stable gradients. 
The full training objective is given by $ \mathcal{L}_{\text{total}} =
    \lambda_x \mathcal{L}_{\text{diff}}^x +
    \lambda_y \mathcal{L}_{\text{diff}}^y +
    \lambda_p \mathcal{L}_{\text{pred}} +
    \lambda_{\rho}^x \mathcal{L}_{\text{stl}}^x +
    \lambda_{\rho}^y \mathcal{L}_{\text{stl}}^y,$
where the coefficients $\lambda_{(\cdot)}$ balance reconstruction quality, forecasting accuracy, and specification compliance across the two stages.

\color{black}
\section{Evaluation}

\subsection{Experimental Setup}
\label{sec:eval-setup}

\paragraph{Datasets.}
We evaluate \textsc{LogiC-Diff} on two CPS domains. Our primary dataset is a \emph{smart city} corpus constructed from the Caltrans PeMS system~\citep{chen2002freeway} for the Los Angeles region. It contains hourly traffic features, including total flow, average occupancy, and average speed, augmented with hourly air-quality and emissions measurements ($\mathrm{NO}_x$, $\mathrm{NO}_2$, ozone, and PMx) from the California Air Resources Board~\citep{carb_air_quality_emissions_data}. The data span 2024--2025 and contain $d{=}7$ features. Each sample uses $m{=}24$ hours of history to forecast the next $H{=}6$ hours. We also consider the San Francisco region of PeMS to compare with the benchmark used by~\cite{liurobust} for sparse indirect attacks. Our second domain is \emph{UAV flight telemetry} from the ALFA dataset~\citep{keipour2021alfa}, which contains 47 fixed-wing flights with real mid-flight engine and actuator faults. We use seven channels that capture the closed-loop dynamics, including GPS groundspeed, altitude, measured and commanded roll and pitch, and throttle output. 

\paragraph{Forecaster and metrics.}
We use a multivariate time-series Transformer~\citep{li2019enhancing} as the forecasting backbone (App.~\ref{sec:layers}). We report two metrics: mean squared error (MSE) over the forecast horizon and Sat\%, the average percentage of logic specifications satisfied by each prediction: $\mathrm{Sat}\% = \frac{1}{N_s}\sum_{i=1}^{N_s} \mathbf{1}[\rho_i \geq 0] \times 100$, where $N_s$ is the number of specifications and $\rho_i$ is the robustness value of the $i$-th specification. MSE captures forecasting accuracy, while Sat\% measures how well the forecast preserves the system level properties used to characterize secure behavior. The complete specification set is provided in App.~\ref{sec:apdx-spec}. 

\paragraph{Baselines.}
We compare \textsc{LogiC-Diff} with baselines representing several defense strategies. \textbf{Default} is the undefended forecaster~\citep{vaswani2017attention}. \textbf{Inherent (STL)} is a logic-regularized model based on STLNet~\citep{ma2020stlnet}, which incorporates STL constraints during training but does not perform inference-time repair. \textbf{Adversarial training} includes AT (Madry)~\citep{madry2018towards} and AT (TRADES)~\citep{zhang2019theoretically}. \textbf{Adversarial purification} is represented by AP (DiffPure)~\citep{nie2022diffusion}, which denoises corrupted inputs through reverse diffusion without using system specifications. \textbf{Randomized smoothing} (RS)~\citep{cohen2019certified,liurobust} averages predictions over Gaussian-perturbed inputs. To separate the effect of logic conditioning from direct feasibility enforcement, we also include three \emph{projection-only} variants that apply the STL projection $P_\varphi$ to the model input (pre-hoc), after inference (post-hoc), or at both stages, without diffusion-based repair or logic conditioning. 

\begin{table*}[t]
\centering
\caption{\textit{Robust} performance under \textit{physical} attacks. MSE and STL Sat\% reported as $\mathrm{mean}_{\pm 95\%~\mathrm{CI}}$.}
\label{tab:physical_attacks}
\small
\resizebox{\textwidth}{!}{%
\begin{tabular}{l|cc|cc|cc|cc}
\toprule
  & \multicolumn{2}{c|}{\textbf{Dropout}}
  & \multicolumn{2}{c|}{\textbf{FDI}}
  & \multicolumn{2}{c|}{\textbf{Replay}}
  & \multicolumn{2}{c}{\textbf{Mixed}} \\
\cmidrule(lr){2-3}\cmidrule(lr){4-5}\cmidrule(lr){6-7}\cmidrule(lr){8-9}
\textbf{Method}
  & MSE$\downarrow$ & Sat\%$\uparrow$
  & MSE$\downarrow$ & Sat\%$\uparrow$
  & MSE$\downarrow$ & Sat\%$\uparrow$
  & MSE$\downarrow$ & Sat\%$\uparrow$ \\
\midrule
Default
  & $0.469_{\pm.017}$ & $77.3_{\pm3.7}$
  & $0.197_{\pm.008}$ & $84.2_{\pm2.2}$
  & $0.538_{\pm.021}$ & $80.5_{\pm2.8}$
  & $0.283_{\pm.017}$ & $84.5_{\pm2.2}$ \\
Inherent (STL)
  & $0.470_{\pm.017}$ & $77.2_{\pm3.5}$
  & $0.207_{\pm.009}$ & $84.3_{\pm2.2}$
  & $0.551_{\pm.023}$ & $79.6_{\pm2.6}$
  & $0.295_{\pm.019}$ & $84.4_{\pm2.0}$ \\
\midrule
AT (Madry)
  & $0.484_{\pm.015}$ & $78.2_{\pm3.5}$
  & $0.175_{\pm.007}$ & $86.1_{\pm2.4}$
  & $0.459_{\pm.018}$ & $80.7_{\pm2.9}$
  & $0.281_{\pm.016}$ & $84.6_{\pm2.5}$ \\
AT (TRADES)
  & $0.489_{\pm.016}$ & $78.7_{\pm3.7}$
  & $0.205_{\pm.008}$ & $85.1_{\pm2.6}$
  & $0.438_{\pm.016}$ & $80.8_{\pm3.2}$
  & $0.287_{\pm.013}$ & $83.7_{\pm2.7}$ \\
\midrule
AP (DiffPure)
  & $0.473_{\pm.017}$ & $77.4_{\pm3.7}$
  & $0.186_{\pm.008}$ & $84.5_{\pm2.2}$
  & $0.525_{\pm.021}$ & $80.5_{\pm2.8}$
  & $0.276_{\pm.017}$ & $84.3_{\pm2.3}$ \\
\midrule
RS (Cohen)
  & $0.468_{\pm.017}$ & $77.4_{\pm3.7}$
  & $0.201_{\pm.008}$ & $84.3_{\pm2.3}$
  & $0.529_{\pm.021}$ & $80.9_{\pm2.8}$
  & $0.290_{\pm.018}$ & $84.7_{\pm2.2}$ \\
\midrule
\textbf{LogiC-Diff}
  & $\mathbf{0.246_{\pm.010}}$ & $\mathbf{84.8_{\pm2.9}}$
  & $\mathbf{0.162_{\pm.007}}$ & $\mathbf{87.9_{\pm2.2}}$
  & $\mathbf{0.207_{\pm.009}}$ & $\mathbf{88.5_{\pm2.3}}$
  & $\mathbf{0.187_{\pm.008}}$ & $\mathbf{87.6_{\pm2.2}}$ \\
\bottomrule
\end{tabular}%
}
\end{table*}

\paragraph{Threat model and attacks.}
We organize the attacks according to the CPS layer they target and the attacker's level of knowledge. At the \textbf{physical layer}, attacks directly corrupt the sensor stream and do not require access to the forecasting model. We consider four sensor-level attacks, each controlled by an attack-specific scalar budget $\varepsilon$: \textbf{Dropout} ($\varepsilon{=}5$), which zeros a contiguous outage window for each sensor to simulate temporary disconnection; \textbf{False Data Injection} (FDI, $\varepsilon{=}1$), which adds a signed bias of up to $\pm1\sigma$ at each timestep; \textbf{Replay} ($\varepsilon{=}0.5$), which replaces the trailing half of the window with earlier genuine measurements; and \textbf{Mixed} ($\varepsilon{=}2$), which samples one of the three attacks for each input to model uncertainty in the fault type.

At the \textbf{cyber layer}, attacks perturb the model input directly under an $\ell_2$ budget of $\varepsilon{=}1.0$. We consider white-box gradient attacks including PGD~\citep{madry2018towards} with $20$ iterations and relative step size $\alpha{=}0.25\varepsilon$, MIM~\citep{dong2018boosting} with momentum decay $1.0$, CW~\citep{carlini2017towards} with $200$ Adam iterations and learning rate $0.01$, and APGD from AutoAttack~\citep{croce2020reliable}, adapted to forecasting by using MSE as the objective while retaining its adaptive step-size schedule, $100$ iterations, and $5$ restarts. Because \textsc{LogiC-Diff} contains a projection $P_\varphi$, gradient-based attacks against the full pipeline use BPDA~\citep{athalye2018obfuscated} to approximate gradients through the projection. We further consider stronger defense-aware settings. \textbf{Proj.-PGD} assumes white-box access to $P_\varphi$ and alternates loss-ascent steps with re-projection onto the logic-feasible set, while \textbf{BPDA/EOT} averages gradients over $K{=}20$ stochastic forward passes. We consider the query-only \textbf{Square Attack}~\citep{andriushchenko2020square}, which removes gradient access entirely and is evaluated with $273$, $500$, and $1000$ proposal steps (query accounting in App.~\ref{sec:apdx-attacks}). Lastly, the sparse indirect attack of~\citep{liurobust} perturbs correlated series rather than the target series itself.

\paragraph{Clean performance.}
Before considering attacks, we first compare clean forecasting performance. Default achieves $0.154_{\pm.006}$ MSE and $87.5_{\pm2.1}$\% Sat; Inherent (STL) achieves $0.164_{\pm.006}$ / $88.1_{\pm1.9}$; AT (Madry) $0.172_{\pm.006}$ / $87.1_{\pm2.4}$; AT (TRADES) $0.200_{\pm.007}$ / $85.3_{\pm2.6}$; AP (DiffPure) $0.157_{\pm.006}$ / $87.3_{\pm2.1}$; RS $0.153_{\pm.006}$ / $87.7_{\pm2.1}$; and \textsc{LogiC-Diff} $0.155_{\pm.006}$ / $88.1_{\pm.2}$. Thus, \textsc{LogiC-Diff} preserves clean forecasting accuracy while tying for the highest clean Sat\%. In contrast, both adversarial-training baselines incur some loss in clean forecasting accuracy. We next evaluate how these methods behave under attack.

\subsection{Robustness Across Various CPS Attack Models}
\label{sec:eval-surfaces}

We begin by evaluating whether \textsc{LogiC-Diff} improves robustness across the two attack surfaces considered in our CPS setting: physical-layer attacks that corrupt sensor measurements and cyber-layer attacks that directly manipulate the model input. For the cyber setting, we also consider adaptive attackers that explicitly account for the defense.

\paragraph{Physical-layer attacks.}
We first evaluate robustness to four sensor-level attacks in Table~\ref{tab:physical_attacks}. \textsc{LogiC-Diff} achieves the lowest MSE and highest Sat\% across all four settings. The gain is modest under FDI, where adversarial training is already effective, but becomes much larger under Dropout and Replay, reducing MSE from $0.468$ to $0.246$ and from $0.438$ to $0.207$, respectively. The advantage also persists under Mixed attacks, where the corruption type varies across samples. This result suggests that \textsc{LogiC-Diff} is less tied to a specific perturbation model than the competing defenses. Rather than targeting one attack mechanism, the embedded specifications provide a system-level prior over valid CPS behavior that remains useful across heterogeneous forms of sensor corruption. This is particularly helpful for temporally structured attacks, where individual sensor values may appear plausible while the overall trajectory violates expected temporal relationships. 

\begin{table*}[t]
\centering
\caption{\textit{Robust} performance under \textit{cyber} attacks ($\ell_2$, $\varepsilon{=}1.0$).
\textbf{(a)} Gradient-based white-box attacks. MSE and STL Sat\% are reported as $\mathrm{mean}_{\pm 95\%~\mathrm{CI}}$
for the main baselines; projection-based baselines added during the extended evaluation are reported using point estimates.
\textbf{(b)} Defense-aware and query-based attacks (MSE): Proj.-PGD, BPDA/EOT, and the gradient-free Square Attack at $273$, $500$, and $1000$ proposal steps (query accounting in App.~\ref{sec:apdx-attacks}).}
\label{tab:gradient_attacks}
\label{tab:additional_attacks}  
\small

\textbf{(a) Gradient-based white-box attacks}

\vspace{0.3em}

\resizebox{\textwidth}{!}{%
\begin{tabular}{l|cc|cc|cc|cc}
\toprule
  & \multicolumn{2}{c|}{\textbf{PGD}}
  & \multicolumn{2}{c|}{\textbf{MIM}}
  & \multicolumn{2}{c|}{\textbf{CW}}
  & \multicolumn{2}{c}{\textbf{APGD}} \\
\cmidrule(lr){2-3}\cmidrule(lr){4-5}\cmidrule(lr){6-7}\cmidrule(lr){8-9}
\textbf{Method}
  & MSE$\downarrow$ & Sat\%$\uparrow$
  & MSE$\downarrow$ & Sat\%$\uparrow$
  & MSE$\downarrow$ & Sat\%$\uparrow$
  & MSE$\downarrow$ & Sat\%$\uparrow$ \\
\midrule

Default
  & $0.324_{\pm.011}$ & $83.1_{\pm2.4}$
  & $0.320_{\pm.011}$ & $82.9_{\pm2.4}$
  & $0.290_{\pm.012}$ & $84.2_{\pm2.3}$
  & $0.325_{\pm.011}$ & $83.0_{\pm2.4}$ \\

Inherent (STL)
  & $0.316_{\pm.012}$ & $83.4_{\pm2.2}$
  & $0.312_{\pm.012}$ & $83.5_{\pm2.2}$
  & $0.283_{\pm.013}$ & $84.9_{\pm2.2}$
  & $0.317_{\pm.012}$ & $83.4_{\pm2.2}$ \\

\midrule
AT (Madry)
  & $0.226_{\pm.009}$ & $85.3_{\pm2.5}$
  & $0.225_{\pm.009}$ & $85.3_{\pm2.6}$
  & $0.186_{\pm.009}$ & $86.6_{\pm2.4}$
  & $0.226_{\pm.009}$ & $85.3_{\pm2.5}$ \\

AT (TRADES)
  & $0.244_{\pm.009}$ & $84.4_{\pm2.7}$
  & $0.244_{\pm.009}$ & $84.4_{\pm2.7}$
  & $0.210_{\pm.008}$ & $85.1_{\pm2.6}$
  & $0.244_{\pm.009}$ & $84.4_{\pm2.7}$ \\

\midrule
AP (DiffPure)
  & $0.301_{\pm.011}$ & $83.4_{\pm2.4}$
  & $0.297_{\pm.011}$ & $83.4_{\pm2.4}$
  & $0.265_{\pm.012}$ & $85.0_{\pm2.3}$
  & $0.300_{\pm.011}$ & $83.5_{\pm2.4}$ \\

\midrule
RS (Cohen)
  & $0.311_{\pm.011}$ & $83.5_{\pm2.4}$
  & $0.307_{\pm.011}$ & $83.5_{\pm2.4}$
  & $0.278_{\pm.012}$ & $84.7_{\pm2.3}$
  & $0.311_{\pm.011}$ & $83.6_{\pm2.4}$ \\

\midrule
\textbf{LogiC-Diff}
  & $0.197_{\pm.008}$ & $\mathbf{86.9_{\pm2.4}}$
  & $0.197_{\pm.008}$ & $\mathbf{86.9_{\pm2.4}}$
  & $0.179_{\pm.008}$ & $\mathbf{87.6_{\pm2.3}}$
  & $0.197_{\pm.009}$ & $\mathbf{86.9_{\pm2.4}}$ \\

\midrule
\multicolumn{9}{c}{\textit{Projection-based Baselines}} \\
\midrule

Pre-hoc Projection
  & $0.277_{\pm.0.010}$ & $88.0_{\pm2.4}$
  & $0.274_{\pm.0.010}$ & $87.8_{\pm2.4}$
  & $0.239_{\pm.0.010}$ & $88.8_{\pm2.3}$
  & $0.278_{\pm.0.010}$ & $88.0_{\pm2.5}$ \\

Post-hoc Projection
  & $0.225_{\pm.0.011}$ & $\mathbf{100.0}$
  & $0.222_{\pm.0.011}$ & $\mathbf{100.0}$
  & $0.203_{\pm.0.011}$ & $\mathbf{100.0}$
  & $0.228_{\pm.0.011}$ & $\mathbf{100.0}$ \\

Pre-hoc \& Post-hoc
  & $0.221_{\pm.0.011}$ & $\mathbf{100.0}$
  & $0.218_{\pm.0.011}$ & $\mathbf{100.0}$
  & $0.200_{\pm.0.011}$ & $\mathbf{100.0}$
  & $0.224_{\pm.0.011}$ & $\mathbf{100.0}$ \\

\textbf{LogiC-Diff-Post-hoc}
  & $\mathbf{0.173_{\pm.0.008}}$ & $\mathbf{100.0}$
  & $\mathbf{0.173_{\pm.0.008}}$ & $\mathbf{100.0}$
  & $\mathbf{0.162_{\pm.0.008}}$ & $\mathbf{100.0}$
  & $\mathbf{0.174_{\pm.0.008}}$ & $\mathbf{100.0}$ \\

\bottomrule
\end{tabular}%
}

\vspace{0.7em}

\textbf{(b) Defense-aware and query-based attacks}

\vspace{0.3em}

\resizebox{0.72\textwidth}{!}{%
\begin{tabular}{l|ccccc}
\toprule
\textbf{Method}
& \textbf{Proj.-PGD}
& \textbf{BPDA/EOT}
& \textbf{Sq@273}
& \textbf{Sq@500}
& \textbf{Sq@1000} \\
\midrule
Default
& $0.3218_{\pm.011}$ & $0.3551_{\pm.025}$ &  $0.2212_{\pm.0095}$ & $0.2464_{\pm.0099}$ & $0.2613_{\pm.0102}$  \\
AT (Madry)
& $0.2393_{\pm.009}$ & $0.2913_{\pm.019}$ & $0.1935_{\pm.0080}$ & $0.2029_{\pm.0081}$ & $0.2103_{\pm.0084}$ \\
AT (TRADES)
& $0.2304_{\pm.009}$ & $0.2817_{\pm.018}$ & $0.1972_{\pm.0080}$ & $0.2041_{\pm.0080}$ & $0.2092_{\pm.0082}$ \\
AP (DiffPure)
&  $0.2975_{\pm.011}$ & $0.3388_{\pm.024}$ & $0.1934_{\pm.0090}$ & $0.2141_{\pm.0093}$ & $0.2232_{\pm.0095}$ \\
RS (Cohen)
& $0.3092_{\pm.011}$ & $0.3468_{\pm.024}$ &  $0.2003_{\pm.0088}$ & $0.2228_{\pm.0091}$ & $0.2342_{\pm.0094}$ \\
\textbf{LogiC-Diff}
& $\mathbf{0.2005_{\pm.008}}$
& $\mathbf{0.2375_{\pm.017}}$
& $\mathbf{0.1723_{\pm.0077}}$
& $\mathbf{0.1782_{\pm.0078}}$
& $\mathbf{0.1833_{\pm.0079}}$ \\ 
\bottomrule
\end{tabular}
}
\vspace{-1em}
\end{table*}

\paragraph{Cyber-layer attacks.}
We next evaluate $\ell_2$-bounded white-box attacks at $\varepsilon{=}1.0$ in Table~\ref{tab:gradient_attacks}. The main result is that \textsc{LogiC-Diff} consistently improves robustness over adversarial training across different gradient-based attack objectives, while logic regularization or diffusion purification alone provides little protection. Compared with the strongest baseline, AT-Madry, \textsc{LogiC-Diff} reduces MSE from $0.226$ to $0.197$ under PGD, from $0.225$ to $0.197$ under MIM, and from $0.226$ to $0.197$ under APGD, with a smaller improvement from $0.186$ to $0.179$ under CW. The similar gains across PGD, MIM, and APGD suggest that the improvement is not specific to a particular attack optimizer. \textsc{LogiC-Diff} also maintains slightly higher specification satisfaction than AT-Madry ($86.9\%$ vs.\ $85.3\%$ under PGD), suggesting that the reduction in forecasting error also improves specification compliance in practice. 

\paragraph{Projection alone is less effective.}
Table~\ref{tab:gradient_attacks} also shows that directly enforcing STL feasibility is not sufficient to obtain the robustness of \textsc{LogiC-Diff}. Under PGD, applying $P_\varphi$ before inference, after inference, or at both stages yields $0.320$, $0.322$, and $0.319$ MSE, respectively, only marginally different from the undefended model. Applying a hard projection after diffusion-based input repair performs better, but does not compare with the full \textsc{LogiC-Diff} model. This suggests that feasibility is a necessary but insufficient condition for accurate reconstruction. Rather than only projecting predictions into the feasible set, \textsc{LogiC-Diff} uses the specifications to guide training and repair, steering the model toward accurate trajectories within that set and providing benefits beyond minimum feasibility enforcement. 

\paragraph{Adaptive and query-based attacks.}
We further evaluate stronger attackers that explicitly account for the defense (Table~\ref{tab:gradient_attacks}(b)). In particular, we design \textbf{Proj.-PGD}, a defense-aware variant of PGD that incorporates the STL projection $P_\varphi$ into the attack loop by alternating gradient-ascent updates with re-projection onto the logic-feasible set and the $\varepsilon$-ball. \textsc{LogiC-Diff} remains the strongest method under this attack, achieving $0.2005$ MSE compared with $0.2304$ for the best baseline. We additionally strengthen BPDA with EOT over stochastic forward passes and evaluate the Square Attack; \textsc{LogiC-Diff} retains the lowest MSE in both settings. These results show that its advantage persists when the attacker explicitly accounts for the projection mechanism or avoids gradient information altogether. Finally, under the gradient-free \textbf{Square Attack}, \textsc{LogiC-Diff} remains best across all query budgets and degrades only modestly as the budget increases, from $0.1723$ MSE at $273$ proposal steps to $0.1833$ at $1000$.

\begin{table*}[t]
\centering
\caption{
Robustness evaluation following the sparse indirect attack setting of Liu et al.
\textbf{(a)} Sparse indirect attacks on our smart-city dataset under deterministic and probabilistic variants.
\textbf{(b)} Evaluation on a seven-series subset of the traffic benchmark used by Liu et al. under the probabilistic attack.
(a) reports MSE on the target series; (b) reports point-forecast normalized deviation of the target lane. Dataset and metric details are provided in App.~\ref{sec:apdx-data} and App.~\ref{sec:apdx-attacks}. Lower is better for both.
}
\label{tab:liu_attack}
\small

\begin{minipage}[t]{0.57\textwidth}
\centering
\textbf{(a) Sparse indirect attack on Caltrans PeMS (District 7)}

\vspace{0.3em}

\resizebox{\textwidth}{!}{%
\begin{tabular}{l|cc|cc}
\toprule
  & \multicolumn{2}{c|}{\textbf{Deterministic}}
  & \multicolumn{2}{c}{\textbf{Probabilistic}} \\
\cmidrule(lr){2-3}\cmidrule(lr){4-5}

\textbf{Method}
  & $\kappa=2$
  & $\kappa=4$
  & $\kappa=2$
  & $\kappa=4$ \\
\midrule

Default
  & $0.2542_{\pm.016}$
  & $0.2755_{\pm.017}$
  & $0.2523_{\pm.016}$
  & $0.2757_{\pm.017}$ \\

Inherent (STL)
  & $0.2561_{\pm.016}$
  & $0.2745_{\pm.016}$
  & $0.2537_{\pm.016}$
  & $0.2748_{\pm.016}$ \\

\midrule

AT (Madry)
  & $0.1754_{\pm.012}$
  & $0.1829_{\pm.012}$
  & $0.1738_{\pm.012}$
  & $0.1830_{\pm.012}$ \\

AT (TRADES)
  & $0.1750_{\pm.011}$
  & $0.1794_{\pm.011}$
  & $0.1737_{\pm.011}$
  & $0.1795_{\pm.011}$ \\

\midrule

AP (DiffPure)
  & $0.2355_{\pm.016}$
  & $0.2491_{\pm.016}$
  & $0.2341_{\pm.016}$
  & $0.2516_{\pm.017}$ \\

\midrule

RS (Cohen)
  & $0.2397_{\pm.016}$
  & $0.2570_{\pm.016}$
  & $0.2381_{\pm.016}$
  & $0.2575_{\pm.016}$ \\

\midrule

\textbf{LogiC-Diff}
  & $\mathbf{0.1226_{\pm.008}}$
  & $\mathbf{0.1286_{\pm.008}}$
  & $\mathbf{0.1217_{\pm.008}}$
  & $\mathbf{0.1287_{\pm.008}}$ \\

\bottomrule
\end{tabular}%
}
\end{minipage}
\hfill
\begin{minipage}[t]{0.40\textwidth}
\centering
\textbf{(b) Sparse indirect attack on Caltrans PeMS (District 4~\citep{liu2022practical})}

\vspace{0.3em}

\resizebox{\textwidth}{!}{%
\begin{tabular}{c|ccc}
\toprule
\textbf{$\kappa$}
  & \textbf{Default}
  & \textbf{AT (TRADES)}
  & \textbf{LogiC-Diff} \\
\midrule

--
  & $0.1505_{\pm.0110}$
  & $0.1511_{\pm.0092}$
  & $\mathbf{0.1392_{\pm.0096}}$ \\

$1/6$
  & $0.2234_{\pm.0131}$
  & $0.1753_{\pm.0101}$
  & $\mathbf{0.1567_{\pm.0102}}$ \\

$2/6$
  & $0.2336_{\pm.0134}$
  & $0.1852_{\pm.0105}$
  & $\mathbf{0.1644_{\pm.0104}}$ \\

$3/6$
  & $0.2377_{\pm.0136}$
  & $0.1893_{\pm.0107}$
  & $\mathbf{0.1678_{\pm.0106}}$ \\

$5/6$
  & $0.2454_{\pm.0138}$
  & $0.1980_{\pm.0109}$
  & $\mathbf{0.1748_{\pm.0108}}$ \\

$6/6$
  & $0.2476_{\pm.0138}$
  & $0.2003_{\pm.0110}$
  & $\mathbf{0.1770_{\pm.0109}}$ \\

\bottomrule
\end{tabular}%
}
\end{minipage}

\vspace{-1em}
\end{table*}

\paragraph{Sparse indirect attacks.}
Finally, we consider an attacker that cannot modify the target series directly. Following~\cite{liurobust}, the attacker instead perturbs a sparse subset of correlated series and degrades the target forecast through the dependencies learned by the model. This setting is particularly relevant to \textsc{LogiC-Diff}, since the embedded specifications capture relationships across signals rather than treating each series independently. Table~\ref{tab:liu_attack} reports results on both our smart-city dataset and the original benchmark of~\cite{liurobust}. \textsc{LogiC-Diff} achieves the lowest target error in every setting, across both deterministic and probabilistic variants and all sparsity levels, improving over the strongest baseline by $28$--$30\%$ on our dataset and by $11$--$12\%$ on the original benchmark, where it is also the most accurate method on clean data. 

\subsection{Generalizability Across Attack Severity, Access, and CPS Domains}
\label{sec:eval-generalization}
We next test how well this robustness carries beyond that setting along three dimensions: stronger attacks, unseen attack mechanisms, and a different CPS domain. Together, these experiments ask whether the advantage of \textsc{LogiC-Diff} depends on a particular perturbation budget, attack used during training, or dataset.

\begin{figure*}[t]
    \centering
    \includegraphics[width=\linewidth]{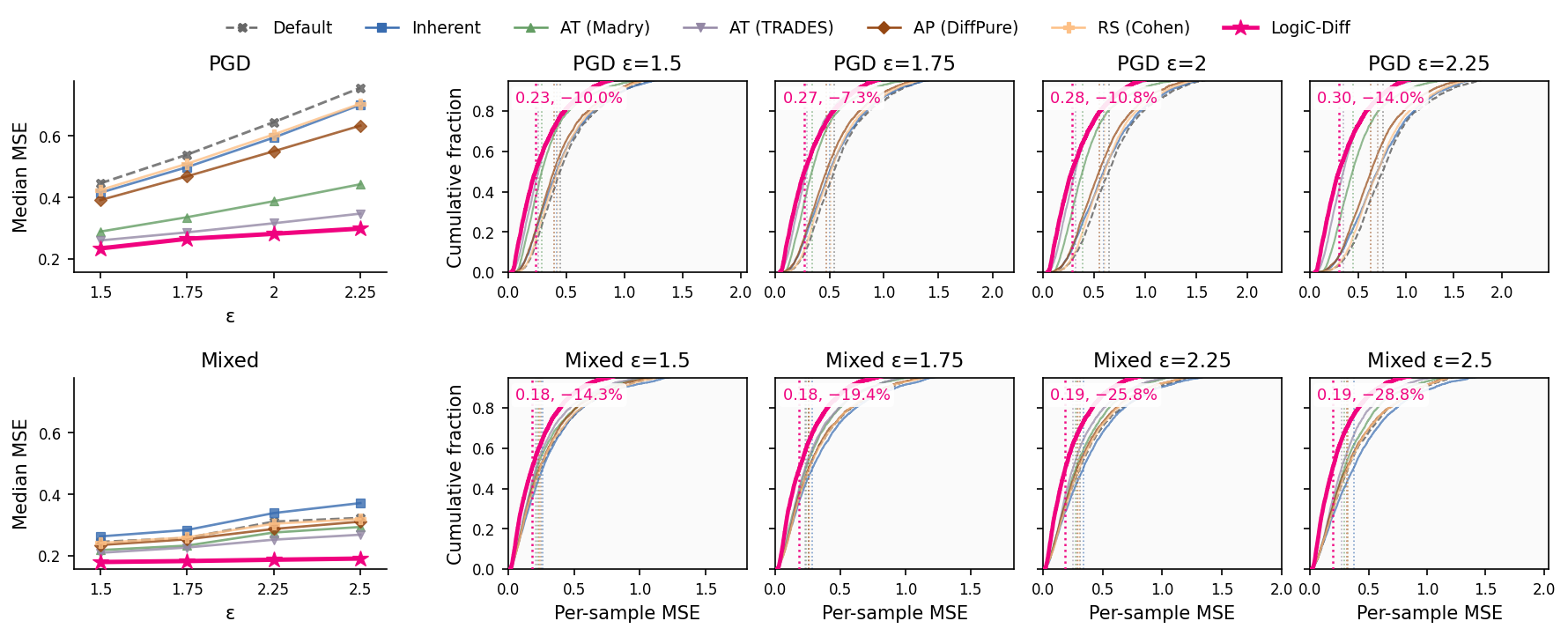}
    \caption{LogiC-Diff achieves the lowest median MSE across all attack budgets and types. \textit{Left}: median per-sample MSE versus attack budget $\varepsilon$ for PGD and Mixed attacks. \textit{Right}: cumulative distributions of per-sample MSE at four budgets per attack family. Annotations show LogiC-Diff's median MSE and the percentage reduction relative to the best-performing baseline at each budget. Vertical dashed lines mark per-method medians. }
    \label{fig:cdf_grid}
    \vspace{-0.75em}
\end{figure*}

\paragraph{Increasing attack strength.}
We first keep the attacks and dataset fixed while increasing the perturbation budget. Figure~\ref{fig:cdf_grid} evaluates PGD with $\varepsilon \in \{1.5, 1.75, 2, 2.25\}$ and Mixed physical attacks with $\varepsilon \in \{1.5, 1.75, 2.25, 2.5\}$. \textsc{LogiC-Diff} achieves the lowest median MSE in all eight settings, outperforming the strongest baseline by $7.3\%$--$28.8\%$. More importantly, its advantage does not disappear as the attack becomes stronger: the margin reaches $14.0\%$ at the largest PGD budget and increases from $14.3\%$ to $28.8\%$ across the Mixed attacks. The effect is especially pronounced for Mixed attacks, where the median MSE of \textsc{LogiC-Diff} remains nearly flat while all baselines degrade. The CDFs show the same pattern across the test distribution, indicating that the improvement is not driven by a small subset of samples. 

\begin{table*}[t]
\centering
\caption{\textit{Generalization} of robustness. \textbf{(a)} Cross-attack generalization: each method is adversarially trained on the opposite attack within the same category. For cyber attacks, the training budget is $\ell_\infty$ with $\varepsilon = 0.1$; for physical attacks, we use the same budgets as in Tables~\ref{tab:physical_attacks} and~\ref{tab:gradient_attacks}. TRADES on MIM-$\ell_\infty$ is omitted as a non-standard configuration. Best per train scenario in \textbf{bold}. Metrics reported as mean $\pm$ 95\% CI.
\textbf{(b)} Cross-domain transfer: \textit{robust} MSE on the ALFA UAV dataset across physical and cyber attacks; lower is better.}
\label{tab:generalization}
\label{tab:cross-attack}  
\label{tab:alfa}  
\small

\textbf{(a) Cross-attack generalization (Caltrans PeMS Smart City)}

\vspace{0.3em}

\resizebox{\textwidth}{!}{%
\begin{tabular}{l|cc|cc|cc|cc}
\toprule
  & \multicolumn{2}{c|}{\shortstack{\textbf{MIM-$\ell_2$}}}
  & \multicolumn{2}{c|}{\shortstack{\textbf{PGD-$\ell_2$}}}
  & \multicolumn{2}{c|}{\shortstack{\textbf{Replay}}}
  & \multicolumn{2}{c}{\shortstack{\textbf{Dropout}}} \\
\cmidrule(lr){2-3}\cmidrule(lr){4-5}\cmidrule(lr){6-7}\cmidrule(lr){8-9}
\textbf{Method}
  & MSE$\downarrow$ & Sat\%$\uparrow$
  & MSE$\downarrow$ & Sat\%$\uparrow$
  & MSE$\downarrow$ & Sat\%$\uparrow$
  & MSE$\downarrow$ & Sat\%$\uparrow$ \\
\midrule
AT (Madry)
  & $0.245_{\pm.009}$ & $84.4_{\pm2.6}$
  & $0.249_{\pm.009}$ & $85.0_{\pm2.5}$
  & $0.582_{\pm.021}$ & $77.0_{\pm3.1}$
  & $0.499_{\pm.019}$ & $76.9_{\pm4.0}$ \\
AT (TRADES)
  & $0.232_{\pm.009}$ & $85.1_{\pm2.5}$
  & --                & --
  & $0.646_{\pm.027}$ & $78.6_{\pm2.8}$
  & $0.550_{\pm.020}$ & $76.5_{\pm3.8}$ \\
\midrule
\textbf{LogiC-Diff}
  & $\mathbf{0.213_{\pm.009}}$ & $\mathbf{87.3_{\pm2.3}}$
  & $\mathbf{0.205_{\pm.009}}$ & $\mathbf{86.5_{\pm2.4}}$
  & $\mathbf{0.275_{\pm.014}}$ & $\mathbf{87.2_{\pm2.2}}$
  & $\mathbf{0.276_{\pm.014}}$ & $\mathbf{85.4_{\pm2.9}}$ \\
\bottomrule
\end{tabular}%
}

\vspace{0.7em}

\textbf{(b) Cross-domain transfer (ALFA UAV)}

\vspace{0.3em}

\resizebox{\textwidth}{!}{%
\begin{tabular}{l|cccc|cccc}
\toprule
  & \multicolumn{4}{c|}{\textit{Physical Attacks}}
  & \multicolumn{4}{c}{\textit{Cyber Attacks}} \\
\cmidrule(lr){2-5}\cmidrule(lr){6-9}
\textbf{Method}
  & \textbf{Dropout}
  & \textbf{FDI}
  & \textbf{Replay}
  & \textbf{Mixed}
  & \textbf{PGD}
  & \textbf{MIM}
  & \textbf{CW}
  & \textbf{APGD} \\
\midrule

Default
  & $0.3036_{\pm.0271}$
  & $0.2351_{\pm.0136}$
  & $1.2283_{\pm.0909}$
  & $0.6246_{\pm.0594}$
  & $0.5036_{\pm.0219}$
  & $0.4876_{\pm.0221}$
  & $0.3476_{\pm.0138}$
  & $0.5350_{\pm.0217}$ \\

Inherent (STL)
  & $0.2958_{\pm.0309}$
  & $0.2523_{\pm.0137}$
  & $1.2774_{\pm.0916}$
  & $0.7305_{\pm.0698}$
  & $0.5086_{\pm.0202}$
  & $0.4867_{\pm.0203}$
  & $0.3496_{\pm.0137}$
  & $0.5362_{\pm.0196}$ \\

\midrule

AT (Madry)
  & $0.3663_{\pm.0334}$
  & $\mathbf{0.1549_{\pm.0107}}$
  & $1.2051_{\pm.0871}$
  & $0.5780_{\pm.0590}$
  & $0.2554_{\pm.0147}$
  & $0.2534_{\pm.0147}$
  & $\mathbf{0.2040_{\pm.0142}}$
  & $0.2560_{\pm.0147}$ \\

AT (TRADES)
  & $0.4304_{\pm.0312}$
  & $0.1727_{\pm.0143}$
  & $1.1387_{\pm.0822}$
  & $0.5542_{\pm.0562}$
  & $0.2637_{\pm.0172}$
  & $0.2627_{\pm.0172}$
  & $0.2213_{\pm.0175}$
  & $0.2639_{\pm.0172}$ \\

\midrule

AP (DiffPure)
  & $0.3457_{\pm.0308}$
  & $0.2140_{\pm.0132}$
  & $1.2045_{\pm.0886}$
  & $0.5890_{\pm.0550}$
  & $0.4358_{\pm.0202}$
  & $0.4203_{\pm.0205}$
  & $0.3111_{\pm.0152}$
  & $0.4525_{\pm.0203}$ \\

\midrule

RS (Cohen)
  & $0.3276_{\pm.0298}$
  & $0.2387_{\pm.0142}$
  & $1.2135_{\pm.0898}$
  & $0.6595_{\pm.0681}$
  & $0.4899_{\pm.0206}$
  & $0.4740_{\pm.0212}$
  & $0.3462_{\pm.0139}$
  & $0.5149_{\pm.0206}$ \\

\midrule

\textbf{LogiC-Diff}
  & $\mathbf{0.1795_{\pm.0175}}$
  & $0.1613_{\pm.0165}$
  & $\mathbf{0.3324_{\pm.0266}}$
  & $\mathbf{0.2197_{\pm.0187}}$
  & $\mathbf{0.2248_{\pm.0191}}$
  & $\mathbf{0.2252_{\pm.0191}}$
  & $0.2063_{\pm.0187}$
  & $\mathbf{0.2243_{\pm.0191}}$ \\

\bottomrule
\end{tabular}%
}
\vspace{-1em}
\end{table*}

\paragraph{Unseen attacks.}
We next ask whether robustness transfers when the attack at test time differs from the one used during adversarial training. Table~\ref{tab:generalization}(a) changes both the attack mechanism and, for cyber attacks, the perturbation norm: models trained on PGD-$\ell_\infty$ are evaluated on MIM-$\ell_2$ and vice versa, while physical models transfer between Dropout and Replay. The main finding is that \textsc{LogiC-Diff} transfers substantially better than conventional adversarial training. On cyber attacks, it reaches $0.213$ MSE on MIM-$\ell_2$ and $0.205$ on PGD-$\ell_2$, compared with $0.232$ and $0.249$ for the strongest AT baselines. The contrast is larger for physical attacks: \textsc{LogiC-Diff} reaches $0.275$ under Replay and $0.276$ under Dropout, reducing MSE by $52.7\%$ and $44.7\%$, respectively, relative to the best AT baseline. In fact, adversarial training on the mismatched physical attack performs worse than the undefended model. This difference reflects an important distinction between the two approaches: adversarial training learns robustness to a particular perturbation distribution, whereas the specifications in \textsc{LogiC-Diff} describe valid system behavior independently of how that behavior is corrupted.

\paragraph{Cross-domain transfer.}
Finally, we evaluate whether the same mechanism remains useful in a physically different CPS. Table~\ref{tab:generalization}(b) repeats the evaluation on the ALFA UAV dataset using the same pipeline and rule families, with specification thresholds re-synthesized from clean flight data. Despite the change in signals, system dynamics, and data scale, \textsc{LogiC-Diff} achieves the lowest MSE in six of the eight attacked settings. The largest gains again occur under temporally structured corruption: MSE decreases from $1.139$ to $0.332$ under Replay, from $0.554$ to $0.220$ under Mixed attacks, and from $0.296$ to $0.180$ under Dropout. Under FDI and CW, the differences from AT-Madry are small and their confidence intervals overlap, showing that the advantage is not uniform across every attack. ALFA also contains roughly three orders of magnitude fewer training windows than the smart-city dataset; for this setting, we use per-sample input-bound calibration. The same qualitative pattern therefore appears in two very different CPS domains: the largest gains occur when attacks disrupt temporal behavior, while additive or optimization-based perturbations leave less room over strong adversarial-training baselines. This provides evidence that the benefit comes from specification-guided repair rather than from characteristics specific to the traffic dataset.

\subsection{Understanding the Sources of Robustness}
\label{sec:eval-sources}

We next examine which parts of \textsc{LogiC-Diff} contribute most to its robustness and how sensitive the framework is to specification quality. Table~\ref{tab:ablation} reports the corresponding ablations.

\begin{table*}[t]
\centering
\caption{Component and logic-rule ablation of LogiC-Diff under Dropout, Mixed, PGD, and CW attacks.
Each variant removes one mechanism or rule family from the full pipeline; metrics are reported as mean $\pm$ 95\% CI.}
\label{tab:ablation}
\small
\resizebox{\textwidth}{!}{%
\begin{tabular}{l|cc|cc|cc|cc}
\toprule
  & \multicolumn{2}{c|}{\textbf{Dropout}}
  & \multicolumn{2}{c|}{\textbf{Mixed}}
  & \multicolumn{2}{c|}{\textbf{PGD}}
  & \multicolumn{2}{c}{\textbf{CW}} \\
\cmidrule(lr){2-3}
\cmidrule(lr){4-5}
\cmidrule(lr){6-7}
\cmidrule(lr){8-9}

\textbf{Variant}
  & MSE$\downarrow$ & Sat\%$\uparrow$
  & MSE$\downarrow$ & Sat\%$\uparrow$
  & MSE$\downarrow$ & Sat\%$\uparrow$
  & MSE$\downarrow$ & Sat\%$\uparrow$ \\
\midrule

-Rule (Input)
  & $0.496_{\pm.020}$ & $79.7_{\pm3.2}$
  & $0.234_{\pm.011}$ & $86.5_{\pm2.2}$
  & $0.250_{\pm.010}$ & $85.8_{\pm2.6}$
  & $0.210_{\pm.010}$ & $87.3_{\pm2.4}$ \\

-Rule (Output)
  & $0.250_{\pm.010}$ & $84.9_{\pm3.0}$
  & $0.194_{\pm.008}$ & $87.2_{\pm2.2}$
  & $0.203_{\pm.008}$ & $86.3_{\pm2.5}$
  & $0.185_{\pm.008}$ & $87.0_{\pm2.4}$ \\

\midrule

-Rule Loss $\mathcal{L}_{\mathrm{stl}}$
  & $0.260_{\pm.010}$ & $84.6_{\pm3.1}$
  & $0.203_{\pm.008}$ & $87.2_{\pm2.3}$
  & $0.219_{\pm.008}$ & $86.4_{\pm2.5}$
  & $0.197_{\pm.008}$ & $87.3_{\pm2.4}$ \\

\midrule

-\texttt{Stability}
  & $0.269_{\pm.012}$ & $85.1_{\pm3.0}$
  & $0.205_{\pm.008}$ & $87.7_{\pm2.2}$
  & $0.224_{\pm.009}$ & $87.1_{\pm2.3}$
  & $0.201_{\pm.009}$ & $87.8_{\pm2.2}$ \\

-\texttt{Associativity}
  & $0.248_{\pm.010}$ & $86.0_{\pm3.1}$
  & $0.193_{\pm.008}$ & $88.0_{\pm2.3}$
  & $0.229_{\pm.008}$ & $87.2_{\pm2.6}$
  & $0.211_{\pm.008}$ & $87.7_{\pm2.5}$ \\

\midrule

\textbf{Full LogiC-Diff}
  & $\mathbf{0.246_{\pm.010}}$ & $84.8_{\pm2.9}$
  & $\mathbf{0.187_{\pm.008}}$ & $87.6_{\pm2.2}$
  & $\mathbf{0.197_{\pm.008}}$ & $86.9_{\pm2.4}$
  & $\mathbf{0.179_{\pm.008}}$ & $87.6_{\pm2.3}$ \\

\bottomrule
\end{tabular}%
}
\vspace{-1em}
\end{table*}

\paragraph{Pipeline components.}
The input-side repair mechanism has the largest effect. Removing it increases Dropout MSE from $0.246$ to $0.496$, Mixed MSE by $25\%$, and PGD MSE by $27\%$. Removing output-stage rule conditioning causes a smaller but consistent degradation, while removing the STL loss also hurts both forecasting accuracy and specification satisfaction. Together, these results suggest that robustness comes primarily from repairing the corrupted input, with training-time logic supervision and output refinement providing additional gains.

\paragraph{Specification families.}
Different rule families contribute to different threats. Removing \texttt{Stability} degrades performance across all attacks, while removing relational rules such as \texttt{Associativity} has a larger effect on PGD and CW. This suggests that envelope constraints help detect implausible individual trajectories, while relational constraints are particularly useful when multiple signals are perturbed jointly or indirectly.

\paragraph{Sensitivity to specification quality.}
The framework is relatively tolerant of loose or incomplete specifications, but much more sensitive to incorrect ones. Widening bounds or removing a moderate fraction of rules has little effect, whereas overly tight, mismatched, or contradictory specifications noticeably degrade performance. Very limited calibration data also hurts, although performance recovers quickly with more calibration samples. Overall, missing guidance is less harmful than misleading guidance.

\paragraph{Discussion.} Our evaluation demonstrates the performance of our method across four settings: fixed-budget physical and cyber attacks, increasing attack budgets, component and rule ablations, and cross-attack generalization under specification guidance, particularly compared with adversarial training baselines. We find that our bi-stage, logic-guided reconstruction and repair mechanism achieves significant performance gains over a wide range of baselines. More importantly, the embedded specifications and overall architecture allow the model to \textit{degrade less drastically under increasing attack strength} and to \textit{generalize more robustly to unseen attacks}. Each rule type contributes meaningfully to overall performance, which indicates that broad non-conflicting specification coverage is usually preferable to relying on any single logic family. 

\paragraph{Limitations.} We further discuss a few potential limitations. First, our evaluation is limited to two important attack families. However, additional threat models, such as training time data poisoning or model extraction would provide more comprehensive security guarantees. Second, STL specifications used by our framework are learned from clean training data and reflect domain assumptions. However, miscalibrated or incomplete specifications could systematically bias predictions in ways that propagate to downstream control. 

\color{black}
\section{Related Work}

\textbf{Adversarial Robustness for Forecasting and CPS.}
Prior research has shown that attacks can occur across the CPS sensing, communication, and learning pipeline~\citep{liurobust,wu2022small,siddiqui2020benchmarking,yu2023survey}, including sensor tampering~\citep{CHEN2023101802}, surrogate-based attacks~\citep{karim2020adversarial,liurobust}, and white-box adaptive attacks~\citep{carlini2017towards}. Existing defenses span CPS-level methods such as detection, resilient estimation, secure control, and fault-tolerant control~\citep{dibaji2019systems}, as well as ML defenses such as adversarial training~\citep{madry2018towards,zhang2019theoretically} and randomized smoothing~\citep{cohen2019certified}. However, CPS deployments may face mixtures of physical and cyber attacks, making single-family defenses insufficient~\citep{duo2022survey}. Our framework handles physical-layer sensor attacks and cyber-layer gradient attacks under a unified defense. 
\textbf{Diffusion-Based Adversarial Defenses and Logic-Aware ML.} Generative defenses repair corrupted inputs by moving them toward the clean data distribution, with diffusion models~\citep{sohl2015deep,ho2020denoising} becoming especially effective. Methods such as Denoised Smoothing~\citep{salman2020denoised}, DiffPure~\citep{nie2022diffusion}, AudioPure~\citep{wu2023defending}, and AAOpt~\citep{wesegoadversary} use diffusion priors or denoising to improve adversarial robustness, but rely mainly on distributional plausibility under $p(x)$ rather than CPS semantics. Separately, logic-aware methods incorporate prior knowledge through training regularization, output filtering, or verification~\citep{scott2020lgml,yang2022logicdef,ma2020stlnet,meng2022adversarial}, and in security settings are often used to constrain training, verify outputs, or filter predictions~\citep{nguyen2024formal,chen2025temporal}. Our work bridges these directions by using logic as semantic guidance within diffusion-based repair, steering reconstruction toward trajectories that are both data-plausible and consistent with CPS security specifications.

\section{Conclusion}
In this paper, we propose a new bi-stage, logic-guided, diffusion-based framework for preventatively enhancing the security of CPS forecasting by embedding formal specifications directly into the prediction pipeline. We identify a broad set of security properties and show how they can be incorporated both before forecasting, through input-side repair, and after forecasting, through output-side refinement. Across more than one hundred specifications in our evaluation, we demonstrate the benefits of the proposed method from multiple perspectives, including forecasting accuracy, specification compliance, robustness under stronger attacks, component and rule ablations, and cross-attack generalization. Overall, the results suggest that the embedded rules provide three main benefits: they improve overall robustness and predictive performance, help the model degrade more gracefully as attack strength increases, and transfer better to unseen attacks. In future work, additional security properties can be formally specified and incorporated within the same framework. Therefore, this work represents an important step toward embedding security properties into CPS forecasting models and suggests that logic-guided repair has strong potential to extend to a broader range of CPS domains.

\bibliography{references}

\clearpage
\newpage
\appendix
\section{General Information}

\paragraph{Code availability.}
Code implementation is available at \url{https://anonymous.4open.science/r/rpc-817A/}.

\paragraph{Computation resouces.} All experiments were conducted on a machine equipped with 128 GB of RAM, an AMD Ryzen Threadripper Pro 7975WX CPU, and an NVIDIA RTX 6000 Ada GPU.

\paragraph{Broader Impact.}
This paper develops methods for improving the security of forecasting models deployed in safety-critical CPS, with direct applications in transportation infrastructure, air-quality monitoring, and autonomous mobility, domains where prediction errors propagate to physical control actions and carry real-world consequences. By embedding formal Signal Temporal Logic specifications directly into a diffusion-based repair pipeline, \textsc{LogiC-Diff} reduces the risk of unsafe downstream behavior under both physical sensor faults (e.g., dropout, replay, false data injection) and adaptive cyber attacks, while degrading more gracefully than purely data-driven defenses as attack strength increases. The framework also generalizes substantially better than adversarial training to attacks not seen during training, which is operationally important for deployments where the threat landscape evolves faster than retraining cycles.

We do not foresee direct negative societal impact from this work, as our contributions are defensive in nature. Two indirect risks merit acknowledgment, however. First, like all logic-guided methods, the framework's behavior depends on the correctness and completeness of the specifications it enforces; miscalibrated or biased specifications could systematically distort predictions in ways that propagate to control decisions, especially in subpopulations underrepresented in the training data used to calibrate feasibility bounds. Second, no defense provides absolute guarantees, and over-reliance on any single robustness mechanism in safety-critical deployments is itself a hazard. We encourage operators to combine logic-guided repair with continued monitoring, redundant sensing, and conventional CPS safety mechanisms.

\section{Notations}
Table~\ref{tab:notation} contains core notations used in this paper. 

\begin{table*}[t]
\centering
\caption{Notations used in this paper.}
\label{tab:notation}
\small
\begin{tabular}{ll}
\toprule
\textbf{Symbol} & \textbf{Description} \\
\midrule
\multicolumn{2}{l}{\emph{Signals and data}} \\
\midrule
$x \in \mathbb{R}^{m \times d}$ & Clean input history window of $d$-dimensional signals over $m$ timesteps. \\
$x^{(j)}_t$ & Value of the $j$-th feature at time $t$. \\
$x^*$ & Adversarially perturbed (attacked) input sequence, $x^* = x + \delta$. \\
$\tilde{x}$ & Repaired input after input-stage logic-conditioned diffusion. \\
$x_n$ & Forward-diffusion latent state at diffusion step $n$, with $n \in \{0, \dots, N\}$. \\
$\hat{x}$ & One-shot $x_0$-estimate from the input-stage reverse diffusion. \\
$x_{\text{logic}}$ & Input-stage logic projection, $x_{\text{logic}} = P_\varphi(x^*)$. \\
$y$ & Ground-truth forecast target. \\
$\hat{y}_{\text{init}}$ & Initial forecast from $F_\omega$ before output-stage refinement. \\
$\hat{y}$ & One-shot $y_0$-estimate from the output-stage reverse diffusion. \\
$y_{\text{logic}}$ & Output-stage logic projection, $y_{\text{logic}} = P_{\varphi'}(\hat{y}_{\text{init}})$. \\
$\tilde{y}$ & Final blended forecast after output-stage refinement. \\
$\delta$ & Adversarial perturbation, with $\|\delta\|_2 \leq \varepsilon$ for $\ell_2$-bounded attacks. \\
\midrule
\multicolumn{2}{l}{\emph{STL specifications}} \\
\midrule
$\varphi$ & Single STL formula. \\
$\varphi$, $\varphi'$ & Input-side and output-side STL specification sets, respectively. \\
$\mu$ & Atomic predicate of the form $f(x_t) \geq c$. \\
$\rho_\varphi(x, t)$ & STL robustness of signal $x$ with respect to $\varphi$ at time $t$. \\
$\tilde{\rho}_\varphi$ & Smooth surrogate of STL robustness used during training. \\
$P_\varphi(\cdot)$ & STL projection operator returning the closest $\varphi$-feasible trace in $\ell_1$ distance. \\
\midrule
\multicolumn{2}{l}{\emph{Models and diffusion process}} \\
\midrule
$F_\omega(\cdot)$ & Forecasting model (transformer-based) with parameters $\omega$. \\
$\Phi_\theta(\cdot)$ & Full end-to-end logic-conditioned diffusion repair pipeline with parameters $\theta$. \\
$\epsilon_\theta(x_n, n, c_x)$ & Input-stage noise predictor conditioned on $c_x$. \\
$\epsilon'_\theta(y_m, m, c_y)$ & Output-stage noise predictor conditioned on $c_y$. \\
$c_x, c_y$ & Logic-derived conditioning signals for input and output stages, respectively. \\
$\beta_n, \alpha_n, \bar{\alpha}_n$ & DDPM noise schedule parameters at diffusion step $n$. \\
$\alpha, \alpha'$ & Learnable blending coefficients for input- and output-stage repair. \\
\midrule
\multicolumn{2}{l}{\emph{Training objective}} \\
\midrule
$\mathcal{L}_{\text{diff}}^x, \mathcal{L}_{\text{diff}}^y$ & Input- and output-side diffusion reconstruction losses. \\
$\mathcal{L}_{\text{stl}}^x, \mathcal{L}_{\text{stl}}^y$ & Input- and output-side smoothed STL feasibility losses. \\
$\mathcal{L}_{\text{pred}}$ & Forecasting loss on the final blended forecast $\tilde{y}$. \\
$\lambda_x, \lambda_y, \lambda_p, \lambda_\rho^x, \lambda_\rho^y$ & Loss weighting coefficients. \\
\bottomrule
\end{tabular}
\end{table*}

\section{Signal Temporal Logic Semantics}\label{sec:stl-semantics}

This section provides additional background on STL, expanding on the inductive definition introduced in Section~\ref{sec:prelim-stl}. We give the full qualitative (Boolean) and quantitative (robustness) semantics~\cite{donze2010robust}, which together support both monitoring and gradient-based optimization over STL formulas.

Throughout this section, $x$ denotes a signal trace, $\mu$ an atomic predicate of the form $f(x_t) \geq c$, and $\varphi$, $\varphi_1$, $\varphi_2$ denote STL formulas. Definition~\ref{def:stlbool} gives the qualitative semantics, where $(x, t) \models \varphi$ reads ``signal $x$ satisfies $\varphi$ at time $t$.''

\begin{definition}[STL Qualitative Semantics]\label{def:stlbool}
\begin{align*}
(x, t) &\models \mu
  && \Leftrightarrow \quad f(x_t) \geq c \\
(x, t) &\models \neg \varphi
  && \Leftrightarrow \quad (x, t) \not\models \varphi \\
(x, t) &\models \varphi_1 \wedge \varphi_2
  && \Leftrightarrow \quad (x, t) \models \varphi_1 \;\text{and}\; (x, t) \models \varphi_2 \\
(x, t) &\models \lozenge_{[a,b]} \varphi
  && \Leftrightarrow \quad \exists t' \in [t+a,\, t+b],\; (x, t') \models \varphi \\
(x, t) &\models \square_{[a,b]} \varphi
  && \Leftrightarrow \quad \forall t' \in [t+a,\, t+b],\; (x, t') \models \varphi \\
(x, t) &\models \varphi_1\, \mathcal{U}_{[a,b]}\, \varphi_2
  && \Leftrightarrow \quad \exists t' \in [t+a,\, t+b],\; (x, t') \models \varphi_2 \\
& && \qquad\quad \text{and}\;\, \forall t'' \in [t,\, t'],\; (x, t'') \models \varphi_1
\end{align*}
\end{definition}

While the qualitative semantics provide a Boolean evaluation of property satisfaction, the quantitative semantics yield a real-valued measure of \emph{how strongly} a signal satisfies or violates a formula. As introduced in Section~\ref{sec:prelim-stl}, the STL robustness function $\rho_\varphi(x, t) \in \mathbb{R}$ measures the signed margin by which $x$ satisfies or violates $\varphi$ at time $t$, with positive values denoting satisfaction and negative values denoting violation. Definition~\ref{def:stlrob} gives the inductive computation.

\begin{definition}[STL Robustness $\rho$]\label{def:stlrob}
\begin{align*}
\rho_\mu(x, t)
  &= f(x_t) - c \\
\rho_{\neg \varphi}(x, t)
  &= -\rho_\varphi(x, t) \\
\rho_{\varphi_1 \wedge \varphi_2}(x, t)
  &= \min\bigl\{\rho_{\varphi_1}(x, t),\; \rho_{\varphi_2}(x, t)\bigr\} \\
\rho_{\lozenge_{[a,b]} \varphi}(x, t)
  &= \max_{t' \in [t+a,\, t+b]}\, \rho_\varphi(x, t') \\
\rho_{\square_{[a,b]} \varphi}(x, t)
  &= \min_{t' \in [t+a,\, t+b]}\, \rho_\varphi(x, t') \\
\rho_{\varphi_1\, \mathcal{U}_{[a,b]}\, \varphi_2}(x, t)
  &= \max_{t' \in [t+a,\, t+b]} \min\Bigl\{\rho_{\varphi_2}(x, t'),\;
       \min_{t'' \in [t,\, t']} \rho_{\varphi_1}(x, t'')\Bigr\}
\end{align*}
\end{definition}

The robustness $\rho_\varphi$ is non-differentiable due to the $\min$ and $\max$ operations. For gradient-based optimization, we use the smooth surrogate $\tilde\rho_\varphi$, which replaces these with differentiable approximations (e.g., soft-min and soft-max) and is the form used in the STL training losses $\mathcal{L}_{\text{stl}}^x$ and $\mathcal{L}_{\text{stl}}^y$.

\section{Background on Physical Attack Models}
We consider adversarial cyber-physical attacks~\cite{duo2022survey} that perturb multivariate spatiotemporal inputs. Following the formulation in Section 2, let $x \in \mathbb{R}^{m \times d}$ denote the clean input window of $d$-dimensional signals over $m$ timesteps, and $x^* = x + \delta$ denote the adversarially perturbed input, where $\delta$ is a perturbation bounded in both pattern and magnitude. Under attack, the forecaster $F_\omega$ produces an erroneous prediction $F_\omega(x^*)$ over the horizon $H$.

The attacker's objective is to manipulate the input in a way that induces violations of task-specific logic constraints, thereby degrading both predictive accuracy and specification compliance. In multivariate settings, the attacker may perturb all features at a given timestamp or selectively target individual ones. Letting $x_t = [x^{(1)}_t, x^{(2)}_t, \dots, x^{(d)}_t]$ denote the $d$-dimensional observation at time $t$, we distinguish two perturbation modes:

\begin{itemize}
  \item \emph{Feature-wise attacks}, which perturb each feature independently,
  $
    x^{*(j)}_t = x^{(j)}_t + \delta^{(j)}_t,
  $
  targeting specific sensor channels (e.g., a compromised speed sensor while flow remains untouched).

  \item \emph{Global attacks}, which apply coordinated perturbations across all features,
  $
    x^*_t = x_t + \delta_t,
  $
  affecting the full observation simultaneously.
\end{itemize}

\section{Dataset Preparation}\label{sec:apdx-data}

We describe the construction of the forecasting windows used in Section~\ref{sec:eval-setup}. All three prepared datasets contain $d=7$ channels, with an input length of $m=24$, a forecast horizon of $H=6$, and a sliding-window stride of one. The saved arrays have shape $(N_s,m,d)$ for inputs and $(N_s,H,d)$ for targets; the evaluation code transposes them to channel-first tensors. Unless otherwise stated, forecasting errors and attack budgets are measured in standardized feature coordinates.

\paragraph{Smart-City Traffic and Air Quality.}
The smart-city data combine hourly Caltrans PeMS District~7 measurements with air-quality observations for the Los Angeles region over 2024--2025~\citep{chen2002freeway,carb_air_quality_emissions_data}. The feature order is total flow, average occupancy, average speed, $\mathrm{NO}_2$, $\mathrm{NO}_x$, ozone, and PM$_{2.5}$; the last channel is denoted PMx in the main text. Traffic stations are matched to the nearest air-quality site within $50$\,km using geographic distance, and records are joined by site and timestamp. In the window-generation script, missing traffic values are filled with station-specific hour-of-day means. Missing air-quality values are forward-filled and backward-filled within station, with station-specific hour-of-day means used for remaining missing entries.

Windows are constructed independently for each station within each monthly source file. The temporal split boundaries are the $70$th and $85$th percentiles of the source timestamps, and each window is assigned according to the timestamp of its final forecast target. The prepared arrays contain $12{,}191{,}904$, $2{,}653{,}074$, and $2{,}650{,}293$ training, validation, and test windows, respectively. The current preprocessing script fits a feature-wise standardizer over all imputed source files before assigning windows to splits. Consequently, this preprocessing uses aggregate statistics beyond the training period; moreover, assigning windows by their final target timestamp does not impose a gap between adjacent splits. These details distinguish the current offline protocol from a strictly training-only, temporally disjoint preprocessing protocol.

\paragraph{UAV Flight Telemetry.}
The ALFA dataset~\citep{keipour2021alfa} contains 47 fixed-wing flight sequences. We use GPS groundspeed, relative altitude, measured and commanded roll, measured and commanded pitch, and throttle PWM output. The asynchronous channels are aligned to a $4$\,Hz grid using the most recent observation at or before each grid time, with a maximum staleness of $1.5$\,s. Rows with an unavailable or stale channel are removed. Windows are formed from consecutive retained rows; their nominal input and output durations are $6$\,s and $1.5$\,s, respectively, although removing invalid rows can introduce temporal gaps.

The earliest recorded failure-status timestamp defines fault onset. A window is classified as clean only when its final target precedes this onset, or when no onset is recorded. Recording sessions, identified by their common flight timestamp, are assigned wholly to one split so that subsequences of the same physical flight cannot appear in different splits. The assignment is deterministic and stratified by fault family, yielding 34/5/8 sequences and $10{,}634$/$1{,}625$/$2{,}907$ clean windows for training/validation/test. Standardization is fitted on the clean training input windows and applied unchanged to targets and other splits. Validation and test windows are shuffled with seed~0 before selecting evaluation subsets. The 840 test windows reaching or following a recorded fault onset are stored separately from the clean-window synthetic-attack evaluation.

\paragraph{Traffic Benchmark for Sparse Indirect Attacks.}
For Table~\ref{tab:liu_attack}(b), we use a seven-series adaptation of the GluonTS \texttt{traffic} dataset used by~\cite{liurobust}, rather than reproducing their complete forecasting setup. The source contains 862 hourly lane-occupancy series, truncated to their common length of $14{,}036$ observations. We select seven correlated series using the first $5{,}000$ hours: start with the series of greatest mean correlation to the others, then greedily add the series of greatest mean correlation to the selected set. The resulting zero-based series indices are $39$, $165$, $669$, $775$, $821$, $834$, and $835$; these identifiers indicate selected correlated series, not a verified spatial ordering of neighboring lanes.

The selected multivariate sequence is split chronologically into $70\%$/$15\%$/$15\%$ segments before windowing. A standardizer fitted on the training segment is reused for validation and test. Windows do not cross segment boundaries, giving $9{,}796$, $2{,}076$, and $2{,}077$ windows, respectively. Validation and test windows are shuffled with seed~0. The sparse attack targets the first selected series, \texttt{lane\_039}, while perturbing a subset of the other six series.

\paragraph{Evaluation Subsets and Specification Calibration.}
The extended attack evaluations use the first $1{,}000$ saved test windows for Proj.-PGD, Square Attack, and each sparse indirect experiment, and the first $150$ for the BPDA/EOT comparison. Thus, attack columns with different subset sizes should not be interpreted as a controlled ranking of attack strength. For these evaluations, the input-side repair bounds are calibrated separately from each unperturbed test input, using two-step windows. Output-side bounds and auxiliary thresholds are pooled from the first $2{,}000$ test target windows. Proj.-PGD additionally constructs a shared attacker-side specification from the first $2{,}000$ clean test inputs. These specifications are held fixed during attack optimization. The per-sample input bounds require a clean reference, and the pooled output calibration uses held-out target information; this is an offline reference-calibrated evaluation, not a training-only specification or an assumption that clean reference signals are available under deployment-time attack.

\section{Attack Implementation and Settings}\label{sec:apdx-attacks}

All attacks operate on input tensors of shape $(B, d, m)$, where $B$ denotes the batch size, $d$ the number of features, and $m$ the input window length. Each attack is parameterized by a scalar $\varepsilon$ that controls attack strength; the precise interpretation of $\varepsilon$ varies by attack family. All attacks preserve tensor shape and are applied independently per batch. We group the attacks into two families:
(i) \emph{black-box physical-layer attacks} that simulate sensor faults and do not require model gradients, and
(ii) \emph{white-box gradient-based attacks} that require gradient access to the evaluated forecasting pipeline. For a defended method, the attacked function includes its inference-time defense, with BPDA used for non-differentiable repair operations. We additionally describe sparse indirect and query-only attacks below.

\paragraph{Physical-Layer Attacks.}
\begin{itemize}
  \item \textbf{False Data Injection (FDI).}
  An $\ell_0$-bounded additive bias attack. For each (sample, sensor) trace, $n_{\text{aff}} = \max\bigl(1, \lceil \min(\varepsilon, 1)\, m \rceil\bigr)$ timesteps are chosen uniformly at random without replacement and perturbed by $s \cdot u$, where $s \in \{-1, +1\}$ is a random sign and $u \sim \mathcal{U}(0, \alpha_{\max})$ with $\alpha_{\max} = \varepsilon \, \hat\sigma(x)$ and $\hat\sigma(x)$ denoting the standard deviation over the input batch. The parameter $\varepsilon$ jointly controls the attacked support (saturating at 100\% of the window for $\varepsilon \geq 1$) and the per-step magnitude ceiling (e.g., $\varepsilon = 1.0$ corresponds to all timesteps perturbed at $\pm 1\sigma$).

  \item \textbf{Temporal Replay.}
  A causal record-and-replay attack. The first $\ell = \mathrm{clip}\bigl(\lceil \varepsilon m \rceil, 1, m/2\bigr)$ timesteps of each (sample, sensor) trace are copied verbatim onto the trailing $\ell$ timesteps; the middle of the sequence is left unmodified. The replay length is capped at $m/2$ to avoid degenerate full-sequence shifts, so $\varepsilon \in (0, 0.5]$ is the meaningful range.

  \item \textbf{Data Dropout.}
  Simulates a sensor disconnect or communication-loss failure. For each (sample, sensor) trace, independently with probability $p_{\text{out}} = \min(1, \varepsilon/3)$, a single contiguous outage of length $\ell_{\text{out}} = \mathrm{clip}\bigl(\lceil 4\varepsilon \rceil, 1, m\bigr)$ is placed at a uniformly random start position, and the affected timesteps are set to zero. The parameter $\varepsilon$ thus controls both the per-sensor failure probability and the outage duration; $\varepsilon \geq 3$ saturates the failure probability so that every sensor fails.

  \item \textbf{Mixed.}
  A randomized ensemble used both at evaluation time and as a physical-layer adversarial-training mode. For each sample in the batch, one attack is drawn uniformly from $\{\text{FDI}, \text{Temporal Replay}, \text{Dropout}\}$ and applied at the shared $\varepsilon$. This setting evaluates robustness under unknown attack-type distributions.
\end{itemize}

\paragraph{Gradient-Based Attacks.}
All four attacks below maximize the per-sample MSE between the forecaster's prediction and the ground-truth horizon, subject to a hard $\ell_p$ budget $\|\delta\|_p \leq \varepsilon$. The target model is held in evaluation mode. Both $\ell_2$ and $\ell_\infty$ variants are provided for PGD, MIM, and APGD; CW is implemented for $\ell_2$ only. Unless stated otherwise, perturbations are initialized uniformly at random inside the $\varepsilon$-ball.

\begin{itemize}
  \item \textbf{PGD.}
  Projected gradient descent on $\delta$ with $20$ iterations and relative step size $\alpha = 0.25\,\varepsilon$. The $\ell_2$ variant uses normalized-gradient steps $\delta \leftarrow \delta + \alpha\, \nabla_\delta \mathcal{L} / \|\nabla_\delta \mathcal{L}\|_2$; the $\ell_\infty$ variant uses sign-gradient steps. After every step, $\delta$ is hard-projected onto the $\varepsilon$-ball.

  \item \textbf{MIM.}
  Momentum Iterative Method. Identical outer loop to PGD, but the gradient at each step is $\ell_1$-normalized and accumulated as an exponential moving average $g_n = \mu\, g_{n-1} + \nabla_\delta \mathcal{L} / \|\nabla_\delta \mathcal{L}\|_1$ with decay $\mu = 1.0$. The step direction is $g_n / \|g_n\|_2$ for $\ell_2$ and $\mathrm{sign}(g_n)$ for $\ell_\infty$. We use $20$ iterations.

  \item \textbf{CW-$\ell_2$.}
  The Carlini--Wagner $\ell_2$ attack: minimize $\|\delta\|_2^2 + c \cdot f(x + \delta, y)$ with Adam (learning rate $0.01$, $c = 10$) for $200$ iterations, where $f$ is the negative regression MSE for the untargeted setting. To remain budget-comparable with PGD and MIM, $\delta$ is hard-projected onto the $\varepsilon$-ball after each Adam step. The best adversarial $\delta$ encountered along the trajectory (largest MSE under the budget, per sample) is returned.

  \item \textbf{APGD.}
  Auto-PGD with adaptive step size, momentum-blended updates
  \[
    x_{n+1} = 0.75 \cdot \mathrm{proj}_\varepsilon(x_n + \alpha_n\, g_n) + 0.25 \cdot (x_n - x_{n-1}),
  \]
  and restart-from-best on plateau detection. The step size is initialized at $2\varepsilon$ and halved whenever the loss fails to improve over a sliding window. We use $100$ iterations and $5$ random restarts; the worst-case (largest-MSE) adversarial example across restarts is reported.
\end{itemize}

\paragraph{BPDA Wrapping for the LogiC-Diff Pipeline.}
The full \textsc{LogiC-Diff} pipeline $\Phi_\theta$ contains a non-differentiable STL projection operator $P_\varphi$. To attack the full pipeline rather than a differentiable surrogate, all gradient-based attacks above are wrapped in the BPDA approximation: the forward pass uses the true $P_\varphi$, while the backward pass treats it as the identity. This is the same protocol used to report the robust-MSE results in Table 3.

\paragraph{EOT Variants for Stochastic Defenses.}
For models with stochastic inference (e.g., diffusion-based purification with noise re-injection), each gradient-based attack has an Expectation-over-Transformations variant that averages the gradient over $K$ independent stochastic forward passes before each PGD, MIM, APGD update. For deterministic models the $K$ gradients are identical and the EOT variant reduces to the base attack.

\paragraph{BPDA/EOT Evaluation Protocol.}
The BPDA/EOT column of Table~\ref{tab:gradient_attacks}(b) uses PGD with an $\ell_2$ budget $\varepsilon=1.0$, $20$ iterations, step size $0.25\varepsilon$, one random initialization, and $K=20$ gradient samples per update. Let $G(x;\xi)$ denote the complete evaluated predictor with inference randomness $\xi$. The attack uses
\[
  g_k = \frac{1}{K}\sum_{r=1}^{K}
  \nabla_{\delta}\mathcal{L}\bigl(G(x+\delta_k;\xi_r),y\bigr),
  \qquad
  \delta_{k+1}=\Pi_{\varepsilon}\left(\delta_k+
  0.25\varepsilon\,\frac{g_k}{\|g_k\|_2}\right),
\]
where $\Pi_{\varepsilon}$ projects each sample's complete perturbation onto the $\ell_2$ ball. Both LogiC-Diff repair hooks use the true repair in the forward pass and an identity backward approximation. When stochastic passes are evaluated as a tiled batch, the per-sample input bounds are repeated in the same order, preserving each replica's conditioning.

LogiC-Diff is evaluated with deterministic one-shot inference and no injected diffusion noise in this comparison; EOT therefore reduces to its base gradient attack. DiffPure uses fresh purification noise with scale $0.1$, while randomized smoothing averages $10$ forecasts with Gaussian input noise of standard deviation $0.25$ per predictor call. Their final errors are averaged over $20$ independent predictor calls, computing the expectation of MSE rather than the MSE of an averaged forecast. The attack initialization is seeded with~0. The comparison uses $150$ test windows, with the calibration protocol in App.~\ref{sec:apdx-data}.

\paragraph{Projection-Aware PGD.}
Proj.-PGD alternates loss ascent with input-side logic repair and perturbation-budget projection. Let $P_{\varphi_g}$ denote the shared attacker-side repair operator, distinct from the per-sample input bounds used by LogiC-Diff. Starting from a random point in the $\ell_2$ ball, each iteration computes
\[
  u_k=\delta_k+0.25\varepsilon\,
       \frac{\nabla_{\delta}\mathcal{L}(G(x+\delta_k),y)}
       {\|\nabla_{\delta}\mathcal{L}(G(x+\delta_k),y)\|_2},
  \qquad
  \delta_{k+1}=\Pi_{\varepsilon}\bigl(P_{\varphi_g}(x+u_k)-x\bigr).
\]
We use $\varepsilon=1.0$, $20$ iterations, one random initialization, and the final iterate. The outer logic repair is applied outside the gradient graph; BPDA remains active inside the defended predictor. Applying the budget projection last enforces the norm constraint, but can undo logic feasibility, so the procedure does not guarantee that the final input belongs to the intersection of both constraint sets. The shared attacker-side specification contains 186 compiled constraints in the saved experiment. Table~\ref{tab:gradient_attacks}(b) uses $1{,}000$ test windows.

\paragraph{Query-Based Square Attack.}
We adapt the score-based random search of~\cite{andriushchenko2020square} to forecasting by maximizing per-sample MSE without using gradients. Each proposal modifies a contiguous block of one feature channel. Initialization assigns an independent random sign to each channel, constant over its input history, then projects the full perturbation onto the $\ell_2$ ball of radius $\varepsilon=1.0$. For a block of length $h$, the proposal replaces its perturbation by a Gaussian vector normalized to length $0.5\varepsilon$, followed by projection of the complete perturbation onto the budget ball. Each sample accepts a proposal only if its queried loss exceeds its incumbent loss. This is a time-series adaptation with budget projection, rather than the original image attack's exact $\ell_2$ budget-transfer procedure.

The block length is $h=\max(1,\mathrm{round}(pm))$, initially $p=0.3$; $p$ is halved after $5\%$, $20\%$, $40\%$, $60\%$, and $80\%$ of the proposal budget. Each budget is evaluated independently with proposal seed~0 and no additional restarts. The saved runs use $273$, $500$, and $1000$ proposal steps on the same $1{,}000$ test windows. There is one additional initialization query and a separate final evaluation call; the proposal budget therefore excludes these two calls.

For DiffPure and randomized smoothing, each query averages $20$ independently sampled predictor outputs before computing the MSE used for acceptance, and final evaluation uses the same averaged predictor. Randomized smoothing itself uses $10$ noisy backbone forecasts per predictor call. Thus, one averaged query requires $20$ predictor calls for DiffPure and $200$ backbone forecasts for randomized smoothing; the table's budgets count proposals, not individual backbone forward passes. Deterministic methods, including the LogiC-Diff configuration above, use one predictor call per query.

\paragraph{Sparse Indirect Attack: Threat Model.}
Following~\cite{liurobust}, the attacker perturbs auxiliary series to corrupt the forecast of protected target series. Let $I$ be the target-channel set and $s(\delta)=|\{j:\|\delta^{(j)}\|_2>0\}|$ count perturbed channels. Our untargeted regression adaptation solves
\[
  \max_{\delta}\;\mathcal{L}_I\bigl(G(x+\delta),y\bigr)
  \quad\text{s.t.}\quad
  \delta^{(j)}=0\ (j\in I),\qquad
  s(\delta)\leq\kappa,\qquad \|\delta\|_2\leq1,
\]
where $\mathcal{L}_I$ is MSE averaged over the forecast horizon and target channels only. Sparsity is imposed over entire feature histories, not individual timestamps. The attack is constructed separately against each evaluated method; LogiC-Diff includes both repair stages with BPDA, and its specifications remain fixed during optimization.

\paragraph{Sparse Indirect Attack: Solvers.}
We implement two support-selection procedures inspired by~\cite{liurobust}:
\begin{itemize}
  \item \textbf{Deterministic.} Initialize a random perturbation inside the $\ell_2$ ball, zero the protected channels, and perform $20$ normalized-gradient ascent steps with step size $0.25$. Retain the $\kappa$ eligible channels of greatest temporal perturbation norm, project back to the budget ball, and refine for $10$ additional steps on this fixed support. Protected channels are zeroed and the norm budget is enforced after every update. One initialization is used, and the final refined perturbation is returned.
  \item \textbf{Probabilistic.} Jointly optimize a raw perturbation $v$ and per-channel logits $\gamma_j$ with Adam for $100$ iterations at learning rate $0.05$. A relaxed mask is sampled as $a_j=\operatorname{sigmoid}((\gamma_j+\log u_j-\log(1-u_j))/\tau)$, with $u_j\sim\mathcal{U}(0,1)$ and $\tau=0.5$. Protected channels are masked out, and $v\odot a$ is projected onto the budget ball before each forward pass. The minimized objective is $-\mathcal{L}_I+10\max(0,\sum_{j\notin I}\operatorname{sigmoid}(\gamma_j)-\kappa)$. Logits start at zero and $v$ starts from Gaussian noise with standard deviation $0.001$. At termination, keep the $\kappa$ eligible channels with the greatest learned probabilities, apply this hard mask to $v$, and project onto the budget ball. The hard mask enforces the final channel budget even though the relaxed optimization uses a soft sparsity penalty.
\end{itemize}
Here, ``probabilistic'' refers to the attack's support-selection mechanism, not to a probabilistic forecasting model. Both solvers maximize target-channel error; they do not use an attacker-specified target trajectory.

\paragraph{Sparse Indirect Attack: Dataset-Specific Settings.}
In Table~\ref{tab:liu_attack}(a), $I$ comprises flow, occupancy, and speed, which remain unmodified. The attacker can perturb at most $\kappa\in\{2,4\}$ of the four air-quality channels. We report target-channel MSE on $1{,}000$ standardized test windows. In Table~\ref{tab:liu_attack}(b), the protected target is \texttt{lane\_039}, and the probabilistic solver can perturb $k\in\{1,2,3,5,6\}$ of the other six selected traffic series. The displayed sparsity fractions are $k/6$; the clean row uses $k=0$. This experiment also uses $1{,}000$ test windows and the same $\ell_2$ budget and probabilistic-solver settings.

For the traffic benchmark, predictions and targets are inverse-standardized before scoring. With $p_{it}$ and $y_{it}$ denoting the target-series prediction and observation in original units, the reported normalized deviation is
\[
  \mathrm{ND}=\frac{\sum_{i=1}^{N_s}\sum_{t=1}^{H}|y_{it}-p_{it}|}
                    {\sum_{i=1}^{N_s}\sum_{t=1}^{H}|y_{it}|}.
\]
This is the point-forecast normalized absolute error, equal to weighted quantile loss at quantile $0.5$; no distinct predictive quantiles are estimated in this adaptation. For Table~\ref{tab:liu_attack}(a), the reported interval is $1.96$ times the standard error of per-window target MSE. For Table~\ref{tab:liu_attack}(b), we resample the $1{,}000$ windows with replacement $2{,}000$ times using seed~0, recompute the ratio of sums, and report $1.96$ times its bootstrap standard deviation. These are window-level intervals and do not account for dependence between overlapping windows.

\begin{table*}[t]
\centering
\caption{Attack types targeting smart-city and cyber-physical predictive pipelines. We focus on adversarial manipulations of inputs to prediction models and organize attacks by increasing operational complexity.}
\label{tab:attacks-cps}
\small
\setlength{\tabcolsep}{6pt}
\renewcommand{\arraystretch}{1.3}
\resizebox{\linewidth}{!}{
\begin{tabular}{@{}>{\raggedright\arraybackslash}p{5.5cm} >{\raggedright\arraybackslash}p{4.0cm} >{\raggedright\arraybackslash}p{2.8cm} >{\raggedright\arraybackslash}p{3.5cm}@{}}
\toprule
\textbf{Attack and Description} & \textbf{Affected Components} & \textbf{Temporal Pattern} & \textbf{Representative Techniques} \\
\midrule
\multicolumn{4}{@{}l}{\emph{Data (Input) Level Attacks}} \\
\midrule
\textbf{Data Suppression.} Suppress or delay genuine events by selectively dropping measurements or interrupting data streams. &
Sensors, gateways, edge aggregators, and time synchronization services. &
Periodic gaps or bursty outages aligned with critical events. &
Selective packet loss, jitter injection, induced clock drift. \\
\addlinespace[2pt]
\textbf{Data Replay.} Reinject previously valid measurements to misrepresent the current system state. &
Sensors and data ingestion pipelines feeding prediction models. &
Event-triggered (e.g., congestion or demand spikes). &
Timestamp-consistent replay, sliding-window or delayed injection. \\
\addlinespace[2pt]
\textbf{False Data Injection.} Inject fabricated or biased sensor values to systematically mislead predictions. &
Loop detectors, CCTV counts, floating-car data, air-quality sensors, communication links. &
Single-shot, periodic, adaptive, or event-driven. &
Value biasing, context-aware perturbations constrained to nominal ranges. \\
\addlinespace[2pt]
\textbf{GNSS Spoofing.} Manipulate physical sensing modalities to distort perceived system state. &
GNSS receivers, Bluetooth and Wi-Fi probes, radar or loop-based traffic sensors. &
Adaptive, often synchronized with traffic peaks. &
RF spoofing, reflective interference, density inflation or deflation. \\
\midrule
\multicolumn{4}{@{}l}{\emph{Model-Level Attacks}} \\
\midrule
\textbf{Sequential Adversarial Examples.} Craft structured perturbations over input sequences to maximize downstream prediction error. &
Sequence predictors (e.g., LSTM, Transformer) and feature normalization layers. &
Inference-time, per prediction window. &
Gradient-based sequence attacks, constraint-aware perturbations. \\
\addlinespace[2pt]
\textbf{Backdoor Poisoning.} Corrupt training data to induce targeted misbehavior when a trigger pattern is present. &
Training pipelines, federated learning clients and aggregators, historical datasets. &
Training-time insertion, with trigger activated at inference. &
Label manipulation, trigger stamping, gradient or parameter perturbation. \\
\bottomrule
\end{tabular}
}
\end{table*}

\section{Framework Implementation Notes}\label{sec:layers}
We describe the full architecture and hyperparameter settings of \textsc{LogiC-Diff} used in all experiments. Following the formulation in Section~3, the pipeline $\Phi_\theta$ consists of three components:
(i) an input-stage diffusion U-Net $\epsilon_\theta$ that purifies the (possibly attacked) input window $x^* \in \mathbb{R}^{B \times m \times d}$ into a repaired signal $\tilde{x}$;
(ii) a sequence-to-sequence forecaster $F_\omega$ that produces an initial horizon prediction $\hat{y}_{\text{init}} \in \mathbb{R}^{B \times H \times d}$;
and (iii) an output-stage diffusion U-Net $\epsilon'_\theta$ that refines the forecast residual.
STL-guided correction is applied at both diffusion stages through the $\ell_1$-projection operator $P_\varphi$ (input side) and $P_{\varphi'}$ (output side), which produce the conditioning signals $c_x$ and $c_y$ (Section~3).
Throughout the experiments we use $d = 7$ features (flow, occupancy, average speed, NOx, NO\textsubscript{2}, ozone, PM\textsubscript{x}), input window length $m = 24$, and forecast horizon $H = 6$.

\paragraph{Diffusion U-Net.}
Both diffusion stages instantiate the same 1D U-Net (DDPM-style $\epsilon$-predictor). When logic conditioning is enabled, the noisy latent and the conditioning signal $c_x$ (or $c_y$) are concatenated along the channel dimension before the input projection.
\begin{itemize}
  \item Input channels: $d = 7$ (or $2d = 14$ when conditioned)
  \item Hidden channels: $C = 32$
  \item Number of levels: $L = 2$, channel multipliers $(1, 2)$
  \item Time embedding: 256-dim sinusoidal $\rightarrow$ MLP
  \item Conv kernel: 3, padding 1; activation: SiLU; normalization: GroupNorm
  \item Down/up sampling: strided convolution (down), nearest-neighbor upsampling (up)
\end{itemize}

The input-stage U-Net operates on temporal length $m = 24$; the output-stage U-Net operates on horizon length $H = 6$. Both follow the standard DDPM encoder--bottleneck--decoder structure with residual blocks and skip connections at every level. The input-stage U-Net (conditioned, $m=24$) maps $(B, 14, 24)$ through an initial Conv1D ($14 \to 32$), two downsampling stages ($32 \to 32$ at length $12$; $32 \to 64$ at length $6$), a residual bottleneck at $(B, 64, 6)$, two upsampling stages with skip connections ($64{+}64 \to 32$; $32{+}32 \to 32$), and a final Conv1D ($32 \to 7$) producing $(B, 7, 24)$. The output-stage U-Net follows the same channel topology applied to the horizon-shaped tensor $(B, 14, 6) \to (B, 7, 6)$.

\paragraph{Diffusion Process.}
We use a discrete-time DDPM formulation:
\begin{itemize}
  \item Number of diffusion steps: $N = 10$ (input stage), $M = 10$ (output stage)
  \item Linear $\beta$ schedule: $\beta_n \in [10^{-4},\, 2\times 10^{-2}]$
  \item Forward: $q(x_n \mid x_0) = \mathcal{N}\!\left(\sqrt{\bar\alpha_n}\, x_0,\; (1 - \bar\alpha_n)\mathbf{I}\right)$
  \item Inference: a single-step $x_0$-estimate at a fixed step $n^\star = N/2 = 5$, recovered from one noise prediction via the closed-form
  \[
    \hat{x} = \frac{x_{n^\star} - \sqrt{1 - \bar\alpha_{n^\star}}\, \epsilon_\theta(x_{n^\star}, n^\star, c_x)}{\sqrt{\bar\alpha_{n^\star}}}.
  \]
  The output stage uses the analogous formula with $m^\star = M/2$ and $\epsilon'_\theta$.
\end{itemize}
The U-Nets are trained with the standard $\epsilon$-prediction objective. The choice of fixed inference step is examined in the ablation study (Section~4), which sweeps $n^\star \in \{1, 3, 5, 7, 9\}$ on the same trained model.

\paragraph{STL-Guided Logic Correction.}
At each diffusion call, the U-Net produces the one-shot estimate $\hat{x}$ (or $\hat{y}$ on the output side), which is passed through the STL projection operator $P_\varphi$ (or $P_{\varphi'}$) to produce the logic-feasible signal $x_{\text{logic}}$ (or $y_{\text{logic}}$). These projected signals serve as both the conditioning input to the diffusion stage and as the reference trajectory for the final blend $\tilde{x} = \alpha \hat{x} + (1 - \alpha) x_{\text{logic}}$. Since $P_\varphi$ is non-differentiable, gradients are propagated through it via straight-through estimation during adversarial-training inner loops, following the BPDA convention~\cite{athalye2018obfuscated}.

\paragraph{Sequence-to-Sequence Forecaster.}
The forecaster $F_\omega$ is an encoder--decoder transformer that consumes the repaired input $\tilde{x} \in \mathbb{R}^{B \times m \times d}$ and produces an initial forecast $\hat{y}_{\text{init}} \in \mathbb{R}^{B \times H \times d}$.
\begin{itemize}
  \item Model dimension: $d_{\text{model}} = 128$
  \item Encoder layers: $3$
  \item Decoder layers: $3$
  \item Attention heads: $4$
  \item Feed-forward dimension: $512$
  \item Dropout: $0.1$
\end{itemize}
Each block uses RMSNorm, multi-head self-attention, optional cross-attention in the decoder, and a SiLU-activated feed-forward sub-block. Sinusoidal positional encodings are added at the encoder and decoder inputs. The decoder is initialized with a learned start token and unrolled auto-regressively over $H$ steps with a causal self-attention mask. Total trainable parameters are approximately $2$M, dominated by the transformer; each U-Net contributes roughly $50$K parameters.

\paragraph{End-to-End Forward Pass.}
For an input $x^*$, the full pipeline $\Phi_\theta$ executes the following steps:
\begin{enumerate}
  \item Input-stage repair: $\tilde{x} \leftarrow \alpha\, \hat{x} + (1 - \alpha)\, x_{\text{logic}}$, where $\hat{x}$ is the U-Net's one-shot estimate and $x_{\text{logic}} = P_\varphi(x^*)$.
  \item Forecasting: $\hat{y}_{\text{init}} \leftarrow F_\omega(\tilde{x})$.
  \item Output-stage refinement: $\tilde{y} \leftarrow \alpha'\, \hat{y} + (1 - \alpha')\, y_{\text{logic}}$, where $\hat{y}$ is the output U-Net's one-shot estimate of $y$ from $\hat{y}_{\text{init}}$ and $y_{\text{logic}} = P_{\varphi'}(\hat{y}_{\text{init}})$.
\end{enumerate}
The blending coefficients $\alpha, \alpha' \in (0, 1)$ are learnable.

\section{Security-Oriented Logic Formulas}\label{sec:apdx-spec}

We formalize security-relevant attack patterns and system consistency constraints using STL, expressed in the same notation as the main-paper templates in Table~\ref{tab:stl-rules}. All formulas are defined over bounded temporal intervals $[t_1,t_2]$ and use the atomic predicate form $x^{(j)}_t \geq c$ from Section~\ref{sec:prelim-stl}, where $x^{(j)}_t$ denotes the value of the $j$-th feature at time $t$. We use the implication shorthand $p \rightarrow q \equiv \neg(p \wedge \neg q)$ for readability; all formulas are equivalent to compositions of operators in our defined grammar.

\paragraph{Adversarial Behavior Models.}
The following formulas characterize the attack patterns in our threat model (Section~2.3). They are not enforced as constraints but used to describe and reason about adversarial signal manipulations.

\begin{itemize}
  \item \textbf{Data suppression.} Periodic suppression of feature values over fixed durations: whenever the sampling phase aligns with the suppression schedule, the feature stays below the operational floor $c_1$ for at least $\tau$ time units, formalized as $\square_{[t_1,t_2]}\bigl((t \equiv 0 \!\!\!\pmod{\Delta}) \rightarrow \square_{[0,\tau]}(\neg(x^{(j)}_{t} \geq c_1))\bigr)$.

  \item \textbf{Replay attack.} Reuse of stale measurements following a triggering condition on feature $x^{(k)}$: when the trigger fires at time $t$, the feature value at $t$ matches its value $\tau$ steps prior within tolerance $c_2$, formalized as $\square_{[t_1,t_2]}\bigl((x^{(k)}_t \geq c_1) \rightarrow \neg(|x^{(j)}_t - x^{(j)}_{t-\tau}| \geq c_2)\bigr)$.

  \item \textbf{False data injection.} Periodic injection of bounded perturbations relative to the clean signal $x^{(j)\ast}_t$: at injection times, the deviation from the clean signal stays below $c_1$, formalized as $\square_{[t_1,t_2]}\bigl((t \equiv 0 \!\!\!\pmod{\Delta}) \rightarrow \neg(|x^{(j)}_t - x^{(j)\ast}_t| \geq c_1)\bigr)$.

  \item \textbf{Sensor spoofing.} Substitution of measurements with crafted values $s^{(j)}_t$ when detection confidence on feature $x^{(k)}$ is low: whenever confidence falls below $c_1$, the feature matches the spoofed signal within $c_2$, formalized as $\square_{[t_1,t_2]}\bigl(\neg(x^{(k)}_t \geq c_1) \rightarrow \neg(|x^{(j)}_t - s^{(j)}_t| \geq c_2)\bigr)$.

  \item \textbf{Adversarial perturbation (inference-time).} Bounded perturbations applied during high-uncertainty windows: when the model uncertainty $u_t$ exceeds $c_1$, the perturbation magnitude stays below $c_2$, formalized as $\square_{[t_1,t_2]}\bigl((u_t \geq c_1) \rightarrow \neg(|x^{(j)}_t - x^{(j)\ast}_t| \geq c_2)\bigr)$.

  \item \textbf{Backdoor poisoning.} Conditional manipulation activated by a trigger feature $p_t$: when the trigger fires, the input matches the manipulated signal $\tilde{x}^{(j)}_t$ within $c_2$, formalized as $\square_{[t_1,t_2]}\bigl((p_t \geq c_1) \rightarrow \neg(|x^{(j)}_t - \tilde{x}^{(j)}_t| \geq c_2)\bigr)$.
\end{itemize}

\paragraph{System Consistency Specifications.}
The following formulas instantiate the rule families from Table~\ref{tab:stl-rules} for specific CPS domains and serve as guidance signals during logic-conditioned repair. Each is annotated with the rule family it belongs to.

\begin{itemize}
  \item \textbf{Physical validity} (\texttt{Stability}). Feature values remain within feasible physical ranges $[c_1, c_2]$: $\square_{[t_1,t_2]}((x^{(j)}_t \geq c_1) \wedge \neg(x^{(j)}_t \geq c_2))$.

  \item \textbf{Capacity constraint} (\texttt{Stability}). Traffic flow cannot exceed physical capacity $c_1$: $\square_{[t_1,t_2]}(\neg(x^{(j)}_t \geq c_1))$.

  \item \textbf{Queue storage limit} (\texttt{Stability}). Queue length stays within storage bound $c_1$: $\square_{[t_1,t_2]}(\neg(x^{(j)}_t \geq c_1))$.

  \item \textbf{Zone-based speed regulation} (\texttt{Stability}, conditional). Speed in zone $z$ respects limit $c_2$ whenever the vehicle is in-zone: $\square_{[t_1,t_2]}((x^{(k)}_t \geq c_1) \rightarrow \neg(x^{(j)}_t \geq c_2))$.

  \item \textbf{Bounded temporal variation} (\texttt{Smoothness}, first order). Consecutive values cannot change abruptly: $\square_{[t_1,t_2]}(\neg(|x^{(j)}_{t+1} - x^{(j)}_t| \geq c_1))$.

  \item \textbf{Travel-time progression} (\texttt{Smoothness}, first order). Travel time changes between adjacent samples stay within $c_1$: $\square_{[t_1,t_2]}(\neg(|x^{(j)}_{t+1} - x^{(j)}_t| \geq c_1))$.

  \item \textbf{Temporal smoothness (second order)} (\texttt{Smoothness}). Acceleration and trend changes remain bounded: $\square_{[t_1,t_2]}(\neg(|x^{(j)}_{t+1} - 2 x^{(j)}_t + x^{(j)}_{t-1}| \geq c_1))$.

  \item \textbf{Cross-sensor consistency} (\texttt{Associativity}, value-level). Co-located sensors agree within tolerance $c_1$: $\square_{[t_1,t_2]}(\neg(|x^{(j)}_t - x^{(k)}_t| \geq c_1))$.

  \item \textbf{Flow conservation} (\texttt{Associativity}, value-level). Inflow and outflow stay balanced within $c_1$: $\square_{[t_1,t_2]}(\neg(|x^{(j)}_t - x^{(k)}_t| \geq c_1))$.

  \item \textbf{Upstream-downstream correlation} (\texttt{Associativity}, value-level). Connected segments remain correlated within $c_1$: $\square_{[t_1,t_2]}(\neg(|x^{(u)}_t - x^{(d)}_t| \geq c_1))$.

  \item \textbf{Spillback constraint} (\texttt{Propagation}). Upstream flow stays bounded by $c_2$ until downstream congestion clears: $\neg(x^{(u)}_t \geq c_2) \, \mathcal{U}_{[t_1,t_2]} \, \neg(x^{(d)}_t \geq c_1)$.

  \item \textbf{Sampling consistency} (\texttt{Liveness}). A valid measurement arrives within every interval of length $\Delta_t$: $\square_{[t_1,t_2]}(\lozenge_{[0,\Delta_t]}(x^{(j)}_t \geq c_1))$.

  \item \textbf{Post-disturbance recovery} (\texttt{Recovery}). After a transient excursion, the feature returns to and remains within a safe band $[c_1, c_2]$ for at least $d$ time units: $\lozenge_{[t_1,t_2]} \, \square_{[0,d]}((x^{(j)}_t \geq c_1) \wedge \neg(x^{(j)}_t \geq c_2))$.
\end{itemize}

\section{Training Configurations and Hyperparameters}
We document the training configuration for all evaluated models, covering optimization, schedules, adversarial-training protocols, and early stopping. The settings below are the script defaults used to reproduce all reported results.

\paragraph{Optimizer and Schedule.}
All models (\textsc{LogiC-Diff} and transformer baselines) are trained with AdamW, initial learning rate $2 \times 10^{-4}$, and weight decay $1 \times 10^{-4}$. The learning rate is held constant throughout training.

\paragraph{Batch Size and Epochs.}
Default batch size is $128$ on CUDA-enabled devices and $64$ on CPU. The default training budget is $50$ epochs on CUDA and $20$ on CPU. More training epochs can be used by setting the corresponding training option.

\paragraph{Loss Weights.}
The total training loss follows Section 3:
\[
  \mathcal{L}_{\text{total}}
  = \lambda_x \mathcal{L}_{\text{diff}}^x
  + \lambda_y \mathcal{L}_{\text{diff}}^y
  + \lambda_p \mathcal{L}_{\text{pred}}
  + \lambda_\rho^x \mathcal{L}_{\text{stl}}^x
  + \lambda_\rho^y \mathcal{L}_{\text{stl}}^y.
\]
Default coefficients: $\lambda_x = \lambda_y = 0.5$, $\lambda_p = 1.0$, $\lambda_\rho^x = \lambda_\rho^y = 0.1$. The two diffusion losses $\mathcal{L}_{\text{diff}}^x, \mathcal{L}_{\text{diff}}^y$ supervise $\hat{x}$ and $\hat{y}$ via the reparameterization trick; the two STL losses $\mathcal{L}_{\text{stl}}^x, \mathcal{L}_{\text{stl}}^y$ apply a softplus penalty to negative robustness $\tilde{\rho}_\varphi(\hat{x}, 0)$ and $\tilde{\rho}_{\varphi'}(\hat{y}, 0)$; and $\mathcal{L}_{\text{pred}}$ is the MSE on the final blended forecast $\tilde{y}$.

\paragraph{Early Stopping.}
We monitor validation MSE and stop training when no improvement is observed for $10$ consecutive epochs. The checkpoint with the lowest validation MSE is used for all reported evaluations.

\paragraph{Adversarial Training.}
The adversarial mode is selected from $\{\text{data-level}, \text{PGD-AT}, \allowbreak \text{MIM-AT}, \allowbreak \text{TRADES}\}$. Default attack injection probability per step is $1.0$ for the gradient-based modes and $0.5$ for the data-level mode. Default inner-loop budgets are $\varepsilon = 0.1$ (PGD/MIM, $\ell_\infty$), $7$ inner steps with relative step size $0.3$, and TRADES $\beta = 6.0$ with $10$ steps. The data-level mode samples from the physical-attack suite described in Section~2 (FDI, replay, dropout, mixed) with the same $\varepsilon$ schedule.

\paragraph{Baselines.}
Transformer-based baselines (AT-Madry, AT-TRADES, AP-DiffPure, undefended) share the same optimizer settings, batch size, epoch budget, and early-stopping policy as \textsc{LogiC-Diff}. The undefended transformer is trained on clean data only; AT-Madry and AT-TRADES use the inner-loop budgets above; AP-DiffPure is trained on clean data and purifies at inference. For the cross-attack generalization experiment (Table~\ref{tab:generalization}(a)), \textsc{LogiC-Diff} is trained with the same adversarial-training recipe as AT methods.

\end{document}